\documentclass[preprint,prd,showpacs,superscriptaddress,preprintnumbers,floatfix]{revtex4}

\usepackage{amsmath}
\usepackage{amssymb}
\usepackage{enumerate}
\usepackage{bm}
\usepackage{dcolumn}
\usepackage{graphicx}
\usepackage{subfigure}

\newenvironment{resulttable}{%
\begin{tabular}{lc@{\hskip-3em}d@{\hskip2em}c@{\hskip1em}c}
\makebox[7em][l]{Integral} & $n_F$ & \makebox[-5em]{Value (Error)}   & Sampling per & No. of \\ [-.5ex]  
                           &       & \makebox[-5em]{including $n_F$} & iteration    & iterations \\ 
\hline
}{%
\end{tabular}
}

\newenvironment{renomtable}{%
\begin{tabular}{l@{\hskip-5em}d@{\hskip2em}l@{\hskip-5em}d}
Integral & \makebox[-4em]{Value(Error)} & 
Integral & \makebox[-4em]{Value(Error)} \\
\hline
}{%
\end{tabular}
}

\begin{document}

\date{\today}

\preprint{RIKEN-TH-202}

\title{%
Tenth-Order QED Contribution to Lepton Anomalous Magnetic Moment $-$
Fourth-Order Vertices Containing Sixth-Order Vacuum-Polarization Subdiagrams
}


\author{Tatsumi Aoyama}
\affiliation{Kobayashi-Maskawa Institute for the Origin of Particles and the Universe (KMI), Nagoya University, Nagoya, 464-8602, Japan}
\affiliation{Theoretical Physics Laboratory, Nishina Center, RIKEN, Wako, Japan 351-0198 }

\author{Masashi Hayakawa}
\affiliation{Theoretical Physics Laboratory, Nishina Center, RIKEN, Wako, Japan 351-0198 }
\affiliation{Department of Physics, Nagoya University, Nagoya, Japan 464-8602 }

\author{Toichiro Kinoshita}
\affiliation{Theoretical Physics Laboratory, Nishina Center, RIKEN, Wako, Japan 351-0198 }
\affiliation{Laboratory for Elementary-Particle Physics, Cornell University, Ithaca, New York, 14853, U.S.A }

\author{Makiko Nio}
\affiliation{Theoretical Physics Laboratory, Nishina Center, RIKEN, Wako, Japan 351-0198 }

\begin{abstract}
This paper reports 
the tenth-order contributions to the $g\!-\!2$ 
of the electron $a_e$ and those of the muon $a_\mu$ from
the gauge-invariant Set~II(c), which consists of 36 Feynman diagrams,
and Set~II(d), which consists of 180 Feynman diagrams.
Both sets are obtained by insertion 
of sixth-order vacuum-polarization diagrams
in the fourth-order anomalous magnetic moment.
The mass-independent contributions from Set~II(c) and Set~II(d) are
     $-0.116~489~(32)(\alpha/\pi)^5$ and 
     $-0.243~00~(29)(\alpha/\pi)^5$,
respectively.
The leading contributions to $a_\mu$, which  involve electron loops only,  
are  $-3.888~27~(90)(\alpha/\pi)^5$ and 
     $0.497~2~(65)(\alpha/\pi)^5$ for Set~II(c) and Set~II(d), respectively.
The total contributions of  
the electron, muon, and tau-lepton loops  to $a_e$ 
are  $-0.116~874~(32) (\alpha/\pi)^5$ for the Set~II(c), 
and  $-0.243~10~(29) (\alpha/\pi)^5$ for the Set~II(d), respectively.
The contributions of electron, muon, and tau-lepton loop to $a_\mu$
are  $-5.559~4 (11) (\alpha/\pi)^5$  for the Set~II(c) 
and  $0.246~5 (65) (\alpha/\pi)^5$ for the Set~II(d), respectively.

\end{abstract}

%

\maketitle

\section{Introduction}
\label{sec:intro}

The anomalous magnetic moment $g\!-\!2$ of the electron has played 
the central role in testing the validity of quantum electrodynamics (QED)
as well as the standard model.
The latest measurement of $a_e\equiv (g\!-\!2)/2$ by the Harvard group 
has reached the precision of $0.24\times 10^{-9}$ 
\cite{Hanneke:2008tm,Hanneke:2010au}:
\begin{eqnarray}
a_e(\text{HV08})= 1~159~652~180.73~ (0.28) \times 10^{-12} ~~~[0.24 \text{ppb}]
~.
\label{a_eHV08}
\end{eqnarray}
At present the best prediction of theory consists of 
QED corrections of up to the eighth order
\cite{Kinoshita:2005sm,Aoyama:2007dv,Aoyama:2007mn}, and
hadronic corrections \cite{Davier:2010nc,Teubner:2010ah,Krause:1996rf,
Melnikov:2003xd,Bijnens:2007pz,Prades:2009tw,Nyffeler:2009tw} 
and electro-weak corrections 
\cite{Czarnecki:1995sz,Knecht:2002hr,Czarnecki:2002nt} 
scaled down from their contributions to the muon $g\!-\!2$.
To compare the theoretical prediction with the experiment 
(\ref{a_eHV08}),
we also need  the value of the fine structure constant $\alpha$
determined by a method independent of $g\!-2\!$ .
The best value of such an $\alpha$ has been obtained recently
from the measurement of $h/m_{\text{Rb}}$, the ratio of the Planck constant
and the mass of Rb atom,  
combined with the very precisely known Rydberg constant and $m_\text{Rb}/m_e$\cite{Bouchendira:2010es}:
\begin{eqnarray}
\alpha^{-1} (\text{Rb10}) = 137.035~999~037~(91)~~~[0.66 \text{ppb}].
\label{alinvRb10}
\end{eqnarray}  
With this  $\alpha$   
the theoretical prediction of $a_e$ becomes 
\begin{eqnarray}
a_e(\text{theory}) = 1~159~652~181.13~(0.11)(0.37)(0.77) \times 10^{-12},
\label{a_etheory}
\end{eqnarray}
where the first, second, and third uncertainties come
from the calculated eighth-order QED term, the tenth-order estimate, and the
fine structure constant (\ref{alinvRb10}), respectively.
The theory (\ref{a_etheory})
is thus in good agreement with the
experiment (\ref{a_eHV08}):  
\begin{eqnarray}
a_e(\text{HV08}) - a_e(\text{theory}) = -0.40~ (0.88) \times 10^{-12},
\end{eqnarray}
proving that QED (standard model) is in good shape even at this very high
precision.

An alternative test of QED is to compare
the $\alpha$ of (\ref{alinvRb10}) with
the value of $\alpha$ determined from the
experiment and theory of $g\!-2\!$~:  
\begin{eqnarray}
\alpha^{-1}(a_e 08) = 137.035~999~085~(12)(37)(33)~~~[0.37 \text{ppb}],
\label{alinvae}
\end{eqnarray}
where the first, second, and third uncertainties come
from the eighth-order QED term, the tenth-order estimate, and the
measurement of $a_e(\text{HV}08)$, respectively.
Although the uncertainty of $\alpha^{-1}(a_e08)$ in (\ref{alinvae}) is a
factor 2 smaller than $\alpha^{-1}(\text{Rb}10)$, it is not a firm
factor since it depends on the estimate of the tenth-order term, which 
is only a crude guess \cite{Mohr:2008fa}.
In anticipating of  this challenge we launched 
a systematic program 
several years ago
to evaluate the complete tenth-order term
\cite{Kinoshita:2004wi,Aoyama:2005kf,Aoyama:2007bs}.

The tenth-order QED contribution to the 
anomalous magnetic moment of an electron can be written as
\begin{equation}
	a_e^{(10)} 
	= \left ( \frac{\alpha}{\pi} \right )^5
         \left [  A_1^{(10)}
	+ A_2^{(10)} (m_e/m_\mu) 
	+ A_2^{(10)} (m_e/m_\tau) 
	+ A_3^{(10)} (m_e/m_\mu, m_e/m_\tau) \right ] ,
\label{eq:ae10th}
\end{equation}
where the electron-muon mass ratio $m_e/m_\mu$  is  $4.836~ 331~ 71~(12)
\times 10^{-3}$
and the electron-tau mass ratio $m_e/m_\tau$ is $2.875~ 64~ (47)\times 10^{-4}$
\cite{Mohr:2008fa}. 
The contribution to the mass-independent term $A_1^{(10)}$ coming from
12672 Feynman diagrams may be
classified into six gauge-invariant sets, further divided into
32 gauge-invariant subsets depending on the nature of closed
lepton loop subdiagrams.
Thus far, results of numerical evaluation of 24 gauge-invariant subsets, 
which consist of 2785 vertex diagrams, 
have been published \cite{Kinoshita:2005sm,Aoyama:2008gy,Aoyama:2008hz,Aoyama:2010yt,Aoyama:2010pk}.
The result of 105 vertex diagrams of Set I(i) has been recently  submitted for publication \cite{Aoyama:2010zp}.
Five of the subsets had also been calculated analytically 
\cite{Laporta:1994md,Aguilar:2008qj}.
Our calculation is in good agreement with these analytic results.

In this article we report the evaluation of contributions of 
two gauge-invariant subsets, Set~II(c) and Set~II(d),
which consist of fourth-order vertex diagrams containing
vacuum-polarization subdiagrams of sixth order.
The effect of insertion of a gauge-invariant set of closed lepton
loops in an internal photon line of momentum $q$ is expressed by
the renormalized vacuum-polarization tensor of the form
\begin{equation}
\Pi^{\mu\nu} (q) = (q^\mu q^\nu -q^2 g^{\mu\nu}) \Pi (q^2),
\end{equation}
where the scalar vacuum-polarization function $\Pi (q^2)$
vanishes at $q^2=0$ on carrying out the charge renormalization.

\begin{figure}
\includegraphics[width=10cm]{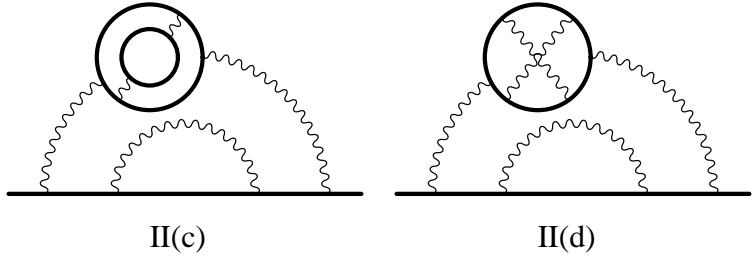}
\caption{Typical diagrams of the tenth-order Set~II(c) and Set~II(d).}
\label{fig:X2cd}
\end{figure}

\begin{figure}
\includegraphics[width=12cm]{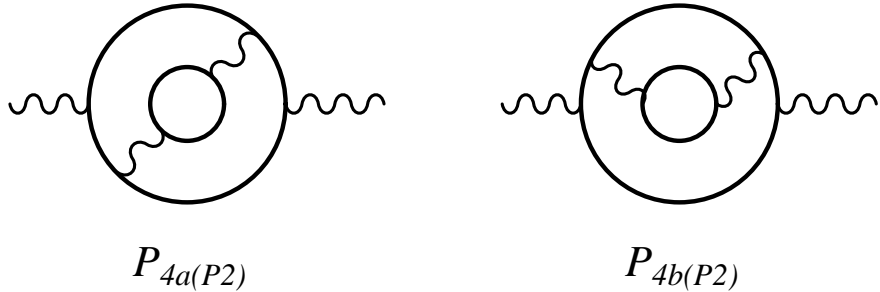}
\caption{Sixth-order vacuum-polarization diagrams consisting of two fermion loops.}
\label{fig:vp4(p2)}
\end{figure}

\begin{figure}
\includegraphics[width=10cm]{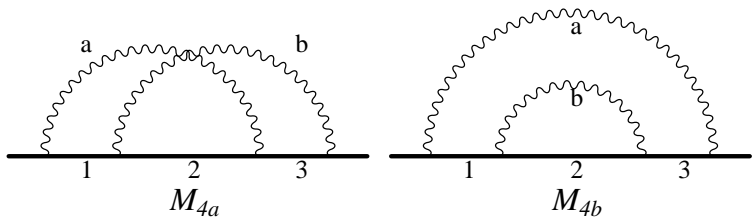}
\caption{Self-energy-like diagrams of fourth order.}
\label{fig:M4ab}
\end{figure}

The Set~II(c) consists of 36 Feynman diagrams.
A typical diagram of this set is shown on the left-hand side of
Fig.~\ref{fig:X2cd}.
It is obtained by insertion 
of proper sixth-order vacuum-polarization 
diagrams containing two closed lepton loops
(see Fig.~\ref{fig:vp4(p2)}) in the fourth-order anomalous magnetic moment 
$M_{4a}$ or $M_{4b}$ represented by Fig.~\ref{fig:M4ab}.
These diagrams can be represented by 4 independent integrals
taking account of various symmetry properties.

The Set~II(d) consists of 180 Feynman diagrams. 
A typical diagram of this set is shown on the right-hand side of
Fig.~\ref{fig:X2cd}.
It is obtained by insertion 
of proper sixth-order vacuum-polarization 
diagrams consisting of one closed lepton loop
in the fourth-order anomalous magnetic moment.
These diagrams can be represented by 16 independent integrals
taking account of various symmetry properties.

Evaluation of the contribution of the Set~II(c) 
to the mass-independent term $A_1^{(10)}$ is straightforward since 
an exact spectral function $\Pi^{(4,2)}$ for the diagrams of
Fig.~\ref{fig:vp4(p2)}  is known for the diagrams 
whose two lepton loops have the same mass \cite{Hoang:1995ex}.
However, evaluation of the mass-dependent term $A_2^{(10)}$ requires $\Pi^{(4,2)}$ as a function of
$m_e/m_\mu$ or $m_e/m_\tau$, which is not available at present.
In order to cover both cases, we follow
an alternative approach \cite{Kinoshita:1990} in 
which we construct a Feynman-parametric
integral of the sixth-order vacuum-polarization function
$\Pi^{(4,2)}$
and insert it in the virtual photon lines of the Feynman-parametric integral of
the forth-order anomalous magnetic moment $M_4$.

\begin{figure}
\includegraphics[width=15cm]{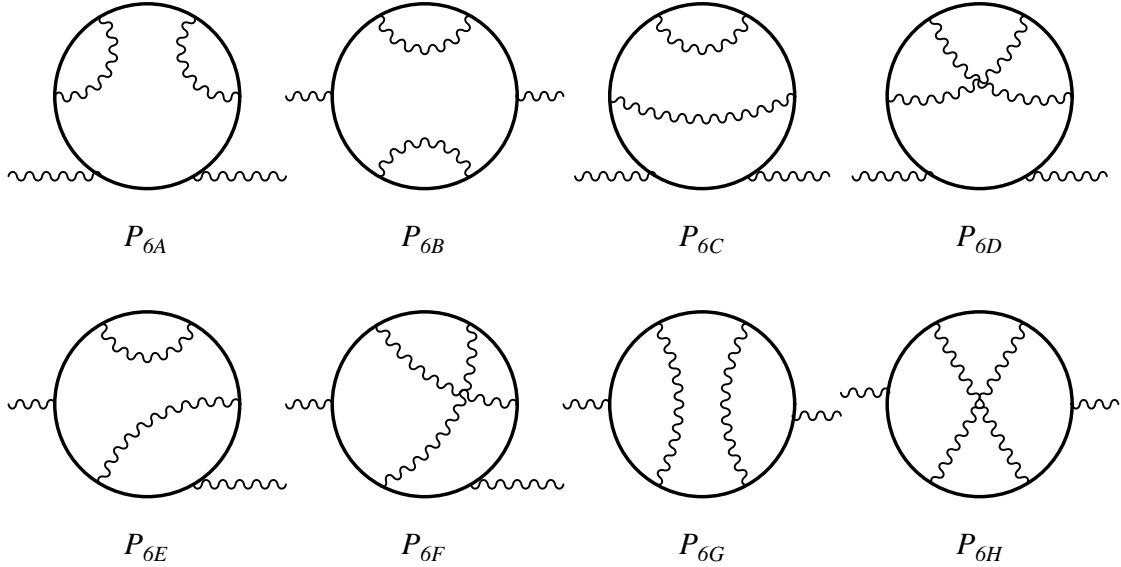}
\caption{Sixth-order vacuum-polarization diagrams consisting of a single fermion loop.}
\label{fig:vp6}
\end{figure}

For the Set~II(d) an exact vacuum-polarization function
$\Pi^{(6)}$ (see Fig.~\ref{fig:vp6}) is not known, although
the Pad\'{e} approximant is known to provide a good approximation \cite{Baikov:1995ui,Kinoshita:1998jg,Kinoshita:1998jf}.
We follow  here primarily the approach \cite{Kinoshita:1990} 
which utilizes the vacuum-polarization function
$\Pi^{(6)}$ itself, instead of its spectral function.
The calculation utilizing the Pad\'{e} approximant of the spectral 
function is also carried out 
to provide an independent check. 

Parametric representations of several vacuum-polarization functions
are presented in Sec.~\ref{sec:vacpol},
where explicit definitions of functions are given.
As an illustration of our approach,
insertion of vacuum-polarization function 
in $M_2$ is presented in Sec.~\ref{sec:vpin2}.
Insertion of $\Pi^{(2)}$, $\Pi^{(4)}$, $\Pi^{(4,2)}$ and $\Pi^{(6)}$ 
in $M_4$ is described in Sec.~\ref{sec:vpin4}.
Although most results of Sec.~\ref{sec:vacpol} and Sec.~\ref{sec:vpin2}
are concerned with quantities of low orders, they are needed
in carrying out the renormalization of the tenth-order terms. 
In cases where the spectral function is available, we present alternative ways
which provide a consistency check of the numerical work.
Application of these methods to Set~II(c) and Set~II(d) is described in
Sec.~\ref{sec:set2c} and Sec.~\ref{sec:set2d}.
Summary and concluding remarks
are presented in Sec.~\ref{sec:summary}.
For simplicity the factor $(\alpha/\pi)^5$ is omitted in Secs. 
\ref{sec:vacpol} -- \ref{sec:set2d}.


\section{Parametric representation of vacuum-polarization function}
\label{sec:vacpol}

As is shown in Ref.~\cite{Kinoshita:1990} the second-order vacuum-polarization
function can be written in the form
\begin{equation}
\Pi^{(2)} (x) = \int (dz) \frac{D_0}{U^2}  \ln \left (\frac{V_0}{V} \right ) ,
\label{pi2}
\end{equation}
with
\begin{eqnarray}
(dz)&=& dz_1 dz_2 \delta(1-z_{12}),~~U=z_{12},~~ V_0=z_{12} m_1^2,~~V=V_0 -x G,
~~x=q^2,
\nonumber  \\
& & ~ G=z_1 A_1,~~ A_1=z_2/U,~~D_0=2 A_1 (1-A_1),
\end{eqnarray}
where $z_1$ and $z_2$ are Feynman parameters
of leptons forming the loop, $z_{12} = z_1 + z_2$,
and $m_1$ is the rest mass of the lepton.

\begin{figure}
\includegraphics[width=12cm]{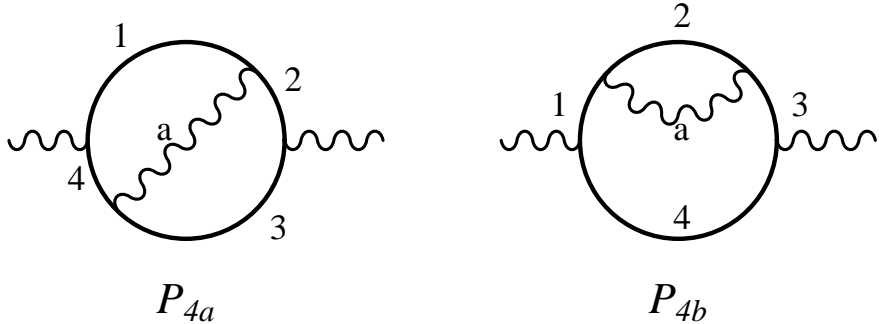}
\caption{Fourth-order vacuum-polarization diagrams
consisting of one lepton loop.}
\label{fig:vp4}
\end{figure}

As a preparation for constructing $\Pi^{(4,2)}$
let us first construct the parametric integral of
the fourth-order vacuum-polarization function $\Pi^{(4)}$.
It has contributions from one diagram $P_{4a}$ and  two diagrams
$P_{4b}$ of Fig.~\ref{fig:vp4}.

The contribution of $P_{4a}$, renormalized at $q=0$ but with subvertex 
divergences not yet removed, can be written as \cite{Kinoshita:1990}
\begin{equation}
\Pi^{(4a)} (x) = \int (dz) \left [ \frac{D_0}{U^2} \left (\frac{1}{V} -\frac{1}{V_0}
\right ) + \frac{D_1}{U^3} \ln \left (\frac{V_0}{V} \right ) \right],
\label{pi4a}
\end{equation}
where $z_1, z_2, z_3,z_4$ are Feynman parameters
for the electron lines and $z_a$ is that of the photon line and
\begin{eqnarray}
 (dz) &=& dz_1 dz_2 dz_3 dz_4 dz_a \delta ( 1-z_{1234a}),~~ z_i \geq 0,
\nonumber \\
B_{11} &=& z_{23a},~~B_{12} = z_a,~~ B_{22}=z_{14a},~~U=z_{14} B_{11} + z_{23} B_{12},
\nonumber \\
A_1 &=& (z_3 B_{12}+z_4 B_{11} )/U, ~~A_2 = (z_3 B_{22} + z_4 B_{12})/U,
\nonumber \\
A_3 &=& A_2 -1, ~~A_4=A_1-1,
\nonumber \\
V_0 &=& z_{1234} m_1^2 + \lambda^2 z_a,~~G=z_1 A_1 + z_2 A_2, ~~V=V_0 -x G,~~ x=q^2,
\nonumber \\
D_0 &=& ((A_1+A_4)(A_2+A_3) - A_1 A_4 - A_2 A_3) m_1^2,
\nonumber \\
D_1 &=& (A_1 A_2 + A_3 A_4) B_{12} - A_1 A_4 B_{22} - A_2 A_3 B_{11}.
\label{defp4a}
\end{eqnarray}
Here $z_{23a}=z_2+z_3+z_a$, etc., and
$\lambda$ is the (infinitesimal) photon mass.

This integral contains ultraviolet(UV) divergences arising from the vertex diagrams
\{ 2,3,a\} and \{1,4,a\}, which can be removed by the $K_{23}$-operation
and $K_{14}$-operation \cite{Kinoshita:1990}, respectively.
The renormalized function $\Pi_{ren.}^{(4a)}$ can be written as
\begin{equation}
\Pi_{ren.}^{(4a)} = \Delta \Pi^{(4a)} - 2 L_2^{\rm R} \Pi^{(2)},
\label{p4a}
\end{equation}
where
\begin{equation}
\Delta \Pi^{(4a)} = (1-K_{23} - K_{14}) \Pi^{(4a)} .
\end{equation}
$\Pi^{(2)}$ is the second-order vacuum-polarization function 
given by Eq.~(\ref{pi2}),
and $L_2^{\rm R} \equiv L_2 - L_2^{
\rm UV}$,
where $L_2^{
\rm UV}$ is the UV-divergent part of the second-order 
vertex renormalization
constant $L_2$ defined by the $K$-operation and $L_2^{\rm R}$
is the remainder including an infrared divergent part of $L_2$.

The renormalized vacuum-polarization term
coming from $P_{4b}$ of Fig.~\ref{fig:vp4} 
may be handled similarly by the $K$-operation \cite{Kinoshita:1990}.
It leads to
\begin{equation}
\Pi_{ren.}^{(4b)} = \Delta \Pi^{(4b)} - 2 B_2^{\rm R} \Pi^{(2)},
\label{p4b}
\end{equation}
where $ B_2^{\rm R} = B_2 - B_2^{
\rm UV}$ \cite{Kinoshita:1990} and
\begin{equation}
\Delta \Pi^{(4b)} (x) = \int (dz) (1-K_2) \left [ \frac{D_0}{U^2} \left (\frac{1}{V} -\frac{1}{V_0}
\right ) + \frac{x C_0}{U^2V}
+ \frac{D_1}{U^3} \ln \left (\frac{V_0}{V} \right ) \right],
\label{pi4b}
\end{equation}
with
\begin{eqnarray}
B_{11} &=& z_{2a},~~B_{12} = z_a,~~ B_{22}=z_{134a},~~U=z_{134} B_{11} + z_{2} B_{12},
\nonumber \\
A_1 &=& z_4 B_{11}/U, ~~A_2 = z_4 B_{12}/U,
A_3 = A_1, ~~A_4=A_1-1,
\nonumber \\
G&=& z_{13} A_1 + z_2 A_2, ~~V=V_0 -x G,~~ x=q^2,
\nonumber \\
D_0 &=& (4 A_1-A_2)A_4 m_1^2, ~~C_0 = -A_1^2 A_2 A_4,~~D_1 = A_1 ( A_1 + 3 A_4) B_{12}.
\label{defp4b}
\end{eqnarray}
$V_0$ and $(dz)$ have the same form as in (\ref{defp4a}).

The function $\Pi^{(4,2)}$ is obtained readily from $\Pi^{(4)}$ by
insertion of the spectral representation of
the vacuum-polarization loop 
\begin{equation}
\frac{\Pi(q^2)}{q^2} = \int_0^1 dt \frac{\rho(t)}{-q^2+4 m_2^2 /(1-t^2)},
\label{spectral_rep}
\end{equation}
in the virtual photon line of $\Pi^{(4)}$ carrying momentum $q$. 
This equation can be regarded as a superposition of vector particles of mass
$4m_2^2 /(1-t^2)$, where $m_2$ is the mass of the inserted loop particle.
The net effect is expressed as the replacement
\begin{equation}
\frac{1}{q^2} \longrightarrow \int_0^1 dt \frac{\rho(t)}{q^2-4 m_2^2/(1-t^2)}.
\label{neteffect}
\end{equation}

For the second-order electron loop 
the spectral function is given by \cite{Kinoshita:1990}
\begin{equation}
\rho^{(2)} (t) = \frac{t^2(1-t^2/3)}{1-t^2}.
\end{equation}
From (\ref{pi4a}) and (\ref{neteffect}) we obtain
\begin{equation}
\Delta \Pi^{(4a,2)} (x) = \int_0^1 dt \rho^{(2)} (t) \int (dz) 
(1-K_{23} - K_{14}) \left [ \frac{D_0}{U^2} \left (\frac{1}{V} -\frac{1}{V_0}
\right ) + \frac{D_1}{U^3} \ln \left (\frac{V_0}{V} \right ) \right],
\label{pi4ap2}
\end{equation}
where $x = q^2$, and $D_0$ and $D_1$ are given in Eq.~(\ref{defp4a}).
$V_0$ and $V$ have the form
\begin{equation}
V=V_0 - x G,~~x=q^2, ~~~V_0 = z_{1234} m_1^2 + z_a \frac{4m_2^2}{1-t^2}.
\label{auxpi4bp2}
\end{equation}
Similarly, from (\ref{pi4b}) and (\ref{neteffect}), we obtain
\begin{equation}
\Delta \Pi^{(4b,2)} (x) = \int_0^1 dt \rho^{(2)} (t) \int (dz) (1-K_2) \left [ \frac{D_0}{U^2} \left (\frac{1}{V} -\frac{1}{V_0}
\right ) + \frac{x C_0}{U^2V}
+ \frac{D_1}{U^3} \ln \left (\frac{V_0}{V} \right ) \right],
\label{pi4bp2}
\end{equation}
where $D_0, C_0$ and $D_1$ are given in Eq.~(\ref{defp4b}),
and $V$ and $V_0$ are given by (\ref{auxpi4bp2}).

In the same manner as in (\ref{pi4a}) and (\ref{pi4b})
the sixth-order vacuum-polarization function $\Pi^{(6)} (q^2)$
can be written in the general form \cite{Kinoshita:1990}
\begin{eqnarray}
\Delta \Pi^{(6)} (x) &=& \int (dz) 
\left [ \frac{D_0}{U^2} \left (\frac{1}{V^2} -\frac{1}{V_0^2} \right ) 
+ \frac{x(B_0 + x C_0)}{U^2V^2} \right .
\nonumber \\
&+& \left . \frac{D_1}{U^3} \left ( \frac{1}{V} -\frac{1}{V_0} \right )
+ \frac{xB_1}{U^3V}
+ \frac{D_2}{U^4} \ln \left (\frac{V_0}{V} \right ) \right],
\label{pi6}
\end{eqnarray}
where
\begin{equation}
V_0 = z_{123456} m_1^2,
\label{pi6def}
\end{equation}
and $D_0$ is $m_1^4$ times $D_0$, and
$B_0$ and $D_1$ are $m_1^2$ times $B_0$ and $D_1$,
given in Ref.~\cite{Kinoshita:1979dy}.
For simplicity $K$-operation is not shown explicitly.
For the sixth-order vacuum-polarization diagrams $P_{6C}$ and $P_{6D}$
of Fig.~\ref{fig:vp6}, which contain 4th-order lepton self-energy
subdiagrams, the $K$-operation subtracts only UV-divergent parts
$\delta m_{4a}^{
\rm UV}$ and $\delta m_{4b}^{
\rm UV}$, which are different from
the full mass-renormalization terms 
$\delta m_{4a}$ and $\delta m_{4b}$. 
This causes no problem for the renormalization  of 
vacuum-polarization function which has no infrared divergence.
However, the fully renormalized formula is simpler if the residual
mass-renormalization term 
$\delta m_{4a}^{\rm R} (\equiv \delta m_{4a} - \delta m_{4a}^{
\rm UV})$ 
and $\delta m_{4b}^{\rm R} (\equiv \delta m_{4b} - \delta m_{4b}^{
\rm UV})$
are also subtracted. 
This  subtraction is performed by the $R$-subtraction method
introduced in Ref.~\cite{Aoyama:2007bs}.
We shall therefore include the $R$-subtraction operation
in Eq.~(\ref{pi6}) whenever it is needed.


\section{Insertion of vacuum-polarization function into the second-order anomaly}
\label{sec:vpin2}

Before discussing insertion of vacuum-polarization (VP) diagram
in $M_4$, let us consider insertion in $M_2$.
The Feynman parametric representation of the second-order
magnetic moment can be written in the form \cite{Kinoshita:1990}
\begin{equation}
M_2 =  \int \frac{(dy)}{U_y^2}  \frac{F_0}{V_y},
\end{equation}
where $y_1$ and $y_a$ are Feynman parameters of the electron line and the photon line,
respectively, 
\begin{eqnarray}
(dy) &=& dy_1 dy_a \delta(1-y_{1a}),~~ y_{1a} = y_1 + y_a, 
\nonumber \\
U_y &=& y_{1a} B_{11},~~ B_{11} = 1,~~ A_1 = y_a/U_y,~~ V_y=y_1 - y_1 A_1, ~~ F_0 = y_1 A_1 (1-A_1),
\end{eqnarray}
and the rest mass of the open electron line (or, open lepton line) is chosen to be 1.
The insertion of the vacuum-polarization function $\Pi$
into the photon line $a$ can be written as
\begin{equation}
M_{2,P} = \int_0^1 dt \rho (t) \int \frac{(dy)}{U_y^2}
\frac{F_0}{V_y+ 4 m_1^2 y_a (1-t^2)^{-1}},
\label{P_insertion}
\end{equation}
following Eq.~(\ref{neteffect}).
Comparing (\ref{P_insertion}) with (\ref{spectral_rep}), we can write 
\begin{eqnarray}
M_{2,P} &=& \int \frac{(dy)}{U_y^2} \frac{F_0}{y_a} 
\int_0^1 dt \frac{\rho(t)}{V_y/y_a+4m_1^2 (1-t^2)^{-1}} \nonumber \\
&=&\int \frac{(dy)}{U_y^2} \frac{F_0}{y_a} \frac{\Pi(q^2)}{q^2}|_{q^2=-V_y/y_a} \nonumber \\
    &=&-\int \frac{(dy)}{U_y^2} \frac{F_0}{V_y}  \Pi (-V_y/y_a). 
\label{m2p}
\end{eqnarray}
This gives us a simple and transparent recipe for
insertion of the vacuum-polarization function in the photon line $a$ of $M_2$:
\begin{equation}
\frac{1}{V_y} \longrightarrow - \frac{1}{V_y}\Pi (- V_y/y_a).
\label{rule_for_V}
\end{equation}
This is identical with Eq.~(5.8) on page 285 of Ref.~\cite{Kinoshita:1990}
noting that 
\begin{equation}
 V_0=z_{12} m_1^2, ~~V=V_0-xG|_{x=-V_y/y_a},~~W=\frac{V}{V-V_0}.
\end{equation}
Note that this derivation requires only 
the analytic property expressed by the spectral representation.
No actual knowledge of the spectral function is required.

Making use of Eq.~(\ref{rule_for_V}) and the form of
$\Pi^{(2)}(x)$ given by Eq.~(\ref{pi2}), 
$M_{2,P_2}$ can be readily written in the form
\begin{equation}
M_{2,P_2} =  - \int_0^1 dy (1-y) 
 \int (dz) \frac{D_0}{U^2} \ln \left (\frac{V_0}{V} \right),
~~~x=-\frac{V_y}{y_a}=-\frac{y^2}{1-y},
\label{nospectral1}
\end{equation} 
where $y=y_1$, $y_a = 1-y_1$, and $V=V_0-xG$.

In the case $P=P_4$, 
we obtain from Eq.~(\ref{pi4b}) 
\begin{equation}
M_{2,P_4} =  - \int_0^1 dy (1-y) 
\int (dz) \left [ \frac{D_0}{U^2} \left (\frac{1}{V} -\frac{1}{V_0} \right ) 
+ \frac{x C_0}{U^2V}
+ \frac{D_1}{U^3} \ln \left (\frac{V_0}{V} \right ) \right]_{x=-y^2/(1-y)},
\label{nospectral2}
\end{equation} 
where  $V_0 = z_{1234} m_1^2$ and $V=V_0-xG$.
$K$-operation (and $R$-subtraction) are not shown explicitly for simplicity.
This corresponds to Eq.~(5.20) in page 288 of Ref.~\cite{Kinoshita:1990}
noting that 
\begin{equation}
 V_0=z_{12} m_1^2, ~~V=V_0-xG|_{x=-V_y/y_a},~~W=\frac{V}{V-V_0}.
\label{defV}
\end{equation}

Similarly, for insertion of $P=P_6$, we have the general structure 
\begin{eqnarray}
M_{2,P_6} &=&  \int_0^1 dy (1-y) 
\int (dz) \left [\frac{D_0}{U^2} \left ( \frac{1}{V^2}-\frac{1}{V_0^2} \right )
+  \frac{x B_0 +x^2 C_0}{U^2 V^2} \right .
\nonumber \\
&+& \left .  \frac{D_1}{U^3} \left ( \frac{1}{V} - \frac{1}{V_0} \right )
+  \frac{x B_1}{U^3 V}
+ \frac{D_2}{U^4} \ln \left (\frac{V_0}{V} \right ) \right]_{x=-y^2/(1-y)},
\label{nospectral3}
\end{eqnarray} 
where $V_0$ is given in Eq.~(\ref{pi6def}).  This equation, with $m_1^2 =1$,
corresponds to Eq.~(5.32) on page 293 of Ref.~\cite{Kinoshita:1990}
except that it includes $R$-subtractions besides $K$-operations for some diagrams.


\section{Insertion of vacuum-polarization function into a fourth-order magnetic moment}
\label{sec:vpin4}

The fourth-order magnetic moment $M_4$,
which consists of two parts $M_{4a}$ and $M_{4b}$,
has the form \cite{Kinoshita:1990}
\begin{equation}
M_4 = \int (dy)\left [  \frac{E_0+{\tilde C}_0}{U_y^2 V_y} 
+ \frac{N_0+Z_0}{U_y^2 V_y^2} + \frac{N_1+Z_1}{U_y^3 V_y} \right ],
\label{m4}
\end{equation}
where $(dy)=dy_1dy_2dy_3dy_ady_b \delta(1-y_{123ab})$,
$y_1, y_2, y_3$ are Feynman parameters of lepton lines, 
$y_a, y_b$ are Feynman parameters of photon lines,
and
\begin{equation}
V_y=y_{123}+ \lambda^2 y_{ab} - G,~~~ G= y_1A_1+y_2A_2+y_3A_3,~~~ {\rm etc.}
\end{equation}
$E_0, {\tilde C}_0, N_0, Z_0, N_1, Z_1$ are functions of
Feynman parameters defined for $M_{4a}$ and $M_{4b}$, respectively.
Their explicit definitions are given in pages 266 and 267 of 
Ref.~\cite{Kinoshita:1990}.
The {\it tilde} on ${\tilde C}_0$ is introduced here to 
avoid confusion with $C_0$ introduced
in the definition of $\Pi$.
When a VP function is inserted into the photon line $a$ by using its spectral function representation, the denominator $V_y$ is replaced by $V_y + y_a R(z)$.
In the case of the second-order VP function $P_2$, where
$z_1, z_2$ are Feynman parameters of two fermions forming
the vacuum-polarization loop, we have $R(z) = m_1^2 z_{12} /(z_1 z_2)$.

For terms of Eq.~(\ref{m4}) 
proportional to $1/V_y$, we can apply the substitution rule (\ref{rule_for_V})
directly.
For the term proportional to $1/V_y^2$ we may rewrite the denominator using the formula
\begin{eqnarray}
  \int (dz) \frac{\rho(z)}{(V_y+y_a R(z))^2} 
 &=&
        - \frac{\partial}{\partial V_y} \int (dz) \frac{\rho(z)}{V_y + y_a R(z)}  
        \nonumber \\
        &=&
          \frac{\partial}{\partial V_y} \frac{\Pi ( -V_y/y_a)}{V_y}
         \nonumber \\ 
        &=&
          - \frac{1}{V_y^2} \Pi(-V_y/y_a)  
          + \frac{1}{V_y} \frac{\partial \Pi(-V_y/y_a)}{\partial V_y}.
\end{eqnarray} 

This leads to the structure of the form
\begin{eqnarray}
M_{4,P} &=&  \int (dy) 
\left [ \left ( \frac{E_0+{\tilde C}_0}{U_y^2 V_y} 
+ \frac{N_0+Z_0}{U_y^2 V_y^2} + \frac{N_1+Z_1}{U_y^3 V_y} \right ) (-\Pi (-V_y/y_a)) 
\right .
\nonumber \\ 
& &\left . + \frac{N_0+Z_0}{U_y^2 V_y} 
\frac{\partial \Pi(-V_y/y_a)}{\partial V_y} \right ].
\label{M4P}
\end{eqnarray}

For $P=P_2$ we have 
\begin{eqnarray}
& & M_{4,P_2} = - \sum_{a,b}  \int (dy) \int (dz) \frac{D_0}{U^2}
\nonumber \\ 
& &\left [  \left ( \frac{E_0+{\tilde C}_0}{U_y^2 V_y} 
+ \frac{N_0+Z_0}{U_y^2 V_y^2} + \frac{N_1+Z_1}{U_y^3 V_y} \right ) \ln \left (
\frac{V_0}{V} \right )
+ \frac{1}{y_a}  \frac{N_0+Z_0}{U_y^2 V_y}\frac{G}{V}  \right ],
\end{eqnarray}
which follows from Eq.~(\ref{pi2}) and
\begin{equation}
  \frac{\partial \Pi^{(2)}(-V_y/y_a)}{\partial V_y}=
- \frac{1}{y_a} \int (dz) \frac{D_0}{U^2} \frac{G}{V} ,
\end{equation}
where $V_0$ and $V$ are given by Eq.~(\ref{defV}),
and $\sum_{a,b}$ means the sum of insertions of $P_2$ in photon lines $a$ and $b$.
Note that the structure of $M_4$ is largely kept intact by the insertion
of $\Pi$.

For $P=P_4$, where $P_4$ represents $P_{4a}$ or $P_{4b}$,
$\Pi^{(4)} (-V_y/y_a)$ is given by Eq.~(\ref{pi4b}) and
its derivative can be written in the form
\begin{equation}
 \frac{\partial \Pi^{(4)}(-V_y/y_a)}{\partial V_y}
= - \frac{1}{y_a} \int (dz)
\left [\frac{D_0}{U^2} \frac{G}{V^2} 
+  \frac{C_0}{ U^2} \left (\frac{1}{V} + \frac{xG}{V^2}\right ) 
+\frac{D_1}{U^3} \frac{G}{V} \right],
\label{pi4_2}
\end{equation}
where, for $P_{4b}$, 
$V=V_0-xG$, $V_0 = z_{1234} m_1^2,~ G=z_{13} A_1 +z_2 A_2,~x=-V_y/y_a $.
Similarly for $P=P_{4a}$.
Substituting Eqs.~(\ref{pi4b}) and (\ref{pi4_2}) in Eq.~(\ref{M4P}), 
we obtain $M_{4a(P_{4b})}$, etc.
To avoid overcrowding the $K$-operation is not shown explicitly.

A formula for $P=P_{4}(P_2)$,
where $P_{4}$ represents $P_{4a}$ or $P_{4b}$, can be readily obtained combining
Eqs.~(\ref{pi4ap2}), (\ref{pi4bp2}), (\ref{M4P}), and (\ref{pi4_2}):
\begin{eqnarray}
M_{4,P_{4}(P_2)} &=&  \int_0^1 dt \rho^{(2)} (t) \int (dy) 
\left [ \left ( \frac{E_0+{\tilde C}_0}{U_y^2 V_y} 
+ \frac{N_0+Z_0}{U_y^2 V_y^2} + \frac{N_1+Z_1}{U_y^3 V_y} \right ) (-\Pi^{(4)} (-V_y/y_a)) 
\right .
\nonumber \\ 
&+&\left . \frac{N_0+Z_0}{U_y^2 V_y}\left (  \frac{\partial \Pi^{(4)} (-V_y/y_a)}{\partial V_y} \right )
 \right ].
\label{M4P4P2}
\end{eqnarray}
where $V$ and $V_0$ are defined by Eq.~(\ref{auxpi4bp2}).

In the case $P=P_6$,
where $P_6$ represents one of $P_{6A},...,P_{6H}$,
we obtain a formula of the form
\begin{eqnarray}
M_{4,P_6} &=& \sum_{a,b} \int (dy) 
\left [ \left ( \frac{E_0+{\tilde C}_0}{U_y^2 V_y} 
+ \frac{N_0+Z_0}{U_y^2 V_y^2} + \frac{N_1+Z_1}{U_y^3 V_y} \right ) (-\Pi^{(6)} (-V_y/y_a)) 
\right .
\nonumber \\ 
&+&\left . \frac{N_0+Z_0}{U_y^2 V_y}
\left (  \frac{\partial \Pi^{(6)} (-V_y/y_a)}{\partial V_y} \right )
 \right ],
\label{M4P6}
\end{eqnarray}
where $\Pi^{(6)} (-V_y/y_a)$ is given by Eq.~(\ref{pi6}) and
its derivative can be written in the form
\begin{eqnarray}
 \frac{\partial \Pi^{(6)}(-V_y/y_a)}{\partial V_y}
&=& - \frac{1}{y_a}    \int (dz)~ \left [
\frac{2D_0}{U^2} \frac{G}{V^3}
+\frac{B_0}{U^2} \left (\frac{1}{V^2} + \frac{2Gx}{V^3} \right )
+ \frac{C_0}{U^2} \left (\frac{2x}{V^2} + \frac{2Gx^2}{V^3}\right )
\right .
\nonumber \\
&+& \left . \frac{D_1}{U^3} \frac{G}{V^2} 
+\frac{B_1}{U^3} \left (\frac{1}{V} + \frac{Gx}{V^2} \right )
+\frac{D_2}{U^4} \frac{G}{V}
\right]_{x=-V_y/y_a}.
\label{pi6_2}
\end{eqnarray}
As usual the $K$-operation and $R$-subtraction are assumed implicitly.
The $K$-operation removes UV-divergent parts of divergent subdiagrams
from $M_{4,P_{6\alpha}}$, $\alpha = A, B,...,H$, etc.
Diagrams $P_{6c}$ and $P_{6d}$ contain fourth-order self-energy subdiagram
so that they require $R$-subtraction in addition to $K$-operation.
The resulting finite quantities are denoted as $\Delta M_{4,P_{6\alpha}}$, etc.
In order to obtain the standard result
renormalized on the mass-shell, further subtraction of UV-finite remainder
must be carried out by the residual renormalization.


\section{Set~II(c)}
\label{sec:set2c}

For the Set~II(c) it is convenient to treat the renormalization of
UV divergences arising from two photons forming $M_{4a}$ and $M_{4b}$
and the renormalization of the vacuum-polarization function separately.
The first step can be written as 
%
\begin{equation}
A_1^{(10)(l_1l_2l_3)}[\text{Set~II(c)}] = \sum_{i=a,b}\Delta M_{4i,P_{4(P_2)}}^{(l_1l_2l_3)}
             - \Delta B_2  M_{2,P_{4(P_2)}}^{(l_1l_2l_3)}  
             - \Delta B_{2, P_{4(P_2)}}^{(l_1l_2l_3)} M_2,
\label{eq:residualSetII(c)}
\end{equation}
where $l_1, l_2$, and $l_3$ denote the open lepton line, outer lepton loop,
and inner lepton loop, respectively.
The superscript $l_1$ is suppressed in $M_2$ and $\Delta B_2$
since they are independent of the lepton mass.
$\Delta B_2$ is the finite part of the second-order renormalization constant 
defined by $ \Delta B_2 \equiv B_2^{\rm R} + L_2^{\rm R} $. (See Eqs.~(\ref{p4a}) and (\ref{p4b}).)
The vacuum-polarization function $P_{4(P_2)}$ 
is fully renormalized whose divergence structure
can be readily found by the $K$-operation.
This leads to the second step:
\begin{align}
&\Delta M_{4i,P_{4(P_2)}}^{(l_1l_2l_3)}
 = \sum_{\beta =a,b} \Delta M_{4i,P_{4\beta (P_2)}}^{(l_1l_2l_3)}
 - 2 \Delta B_{2, P_2}^{(l_2l_3)} \Delta M_{4i,P_2}^{(l_1l_2)},~~~\text{for}~i=a,b,
\label{eq:M4P4(P2)}
\\
&  M_{2,P_{4(P_2)}}^{(l_1l_2l_3)}
        =  \sum_{\beta=a,b}  M_{2 ,P_{4 \beta (P_2)} }^{(l_1l_2l_3)}
           - 2 \Delta B_{2, P_2}^{(l_2l_3)}  M_{2, P_2}^{(l_1l_2)}~,
\label{eq:M2P4(P2)}
\\
& \Delta B_{2,P_{4(P_2)}}^{(l_1l_2l_3)}
        =  \sum_{\beta=a,b} \Delta B_{2, P_{4 \beta (P_2)} }^{(l_1l_2l_3)}
           - 2 \Delta B_{2, P_2}^{(l_2l_3)} \Delta B_{2, P_2}^{(l_1l_2)}.
\label{db2p4p2}
\end{align}
%


\subsection{Numerical results: $(eee)$ case}
\label{subsec:eeeresults}

The contribution of Set~II(c) to the electron $g\!-\!2$
for the case $(l_1l_2l_3)=(eee)$, where $e$ denotes electron,
has been evaluated from Eq.~(\ref{M4P4P2})  by three different methods:

(a) A straightforward extension of the method developed in \cite{Kinoshita:1990},

(b) Method based on the automatic code generating algorithm
{\sc gencodevp}{\it N} \cite{Aoyama:2010zp}, and

(c) Use of an exact spectral function 
of $\Pi^{(4,2)}$ given in Ref.~\cite{Hoang:1995ex}.

\noindent
All calculations are carried out by the integration routine VEGAS \cite{Lepage:1977sw}.
Preliminary evaluations of the integral (\ref{M4P4P2}) 
by the methods (a) and (b) gave results 
consistent with each other within the numerical uncertainty estimated by VEGAS,
proving that both programs are bug-free.
(Actually, both methods (a) and (b) use only $K$-operation since
$R$-subtraction is not needed in this case.) 
We therefore list only the results of method (b) in
the first four data lines of Table~\ref{table:setII(c)_eee}.
The values of auxiliary functions
$\Delta B_{2,P_{4(P_2)}}$ and $ M_{2,P_{4(P_2)}}$ are 
listed in Table~\ref{table:setII(c)residual_eee}.

\renewcommand{\arraystretch}{0.80}
\begin{table}
\begin{ruledtabular}
\caption{ Contributions of diagrams
of Set~II(c) to $a_e$ for $(l_1l_2l_3) = (eee)$.
The superscript $(eee)$ is omitted for simplicity.
The multiplicity $n_F$ is the number of vertex diagrams 
represented by the integral and 
is incorporated in the numerical value. 
The top four lines
are obtained by constructing the sixth-order vacuum-polarization function
in terms of Feynman parameters. The bottom two lines are
obtained by using the exact spectral function of the sixth order.
All integrals are evaluated in double precision. 
\\
\label{table:setII(c)_eee}
}
\begin{resulttable}
$\Delta M_{4a,P_{4a(P_2)}}$&6  &  0.028~927~(21)& $1 \times 10^7,~1\times10^8$ &50,~50 \\
$\Delta M_{4a,P_{4b(P_2)}}$&12 &  0.004~521~(11)& $1 \times 10^7,~1\times10^8$ &50,~50 \\
$\Delta M_{4b,P_{4a(P_2)}}$&6  & -0.110~617~(16)& $1 \times 10^7,~1\times10^8$ &50,~50 \\
$\Delta M_{4b,P_{4b(P_2)}}$&12 & -0.020~212~( 9)& $1 \times 10^7,~1\times10^8$ &50,~50 \\
                            & & & & \\                   
$\Delta M_{4a,P_{4(P_2)}} $&18 &  0.028~425~(28)& $1 \times 10^7,~1\times10^8$ &50,~100 \\
$\Delta M_{4b,P_{4(P_2)}} $&18 & -0.112~236~(20)& $1 \times 10^7,~1\times10^8$ &50,~100 
\end{resulttable}
\end{ruledtabular}
\end{table}
\renewcommand{\arraystretch}{1}


\renewcommand{\arraystretch}{0.70}
\begin{table}
\begin{ruledtabular}
\caption{ Finite renormalization terms of Set~II(c) for the case $(eee)$.
All integrals are evaluated in double precision.
The quantities in the fourth line are obtained by using 
the exact spectral function of the sixth order.
The multiplicity of the integral  
is incorporated in the numerical value. 
\\
\label{table:setII(c)residual_eee}
}
\begin{renomtable}
 $\Delta B_{2,P_2}$             & 0.063~399~266\cdots &
 $ M_{2,P_2}$                   & 0.015~687~421\cdots \\
 $\Delta B_{2,P_{4a(P_2)}}$     & 0.047~836~(1)   &
 $ M_{2,P_{4a(P_2)}}$           & 0.011~403~(1)  \\
 $\Delta B_{2,P_{4b(P_2)}}$     & 0.008~783~(1)   &
 $ M_{2,P_{4b(P_2)}}$           & 0.001~717~(1)  \\
 & & & \\
 $\Delta B_{2,P_{4(P_2)}}$      & 0.048~577~(5)   &
 $ M_{2,P_{4(P_2)}}$            & 0.011~131~(1)  \\
 & & & \\
$ \Delta M_{4a,P_2}$             &  0.039~642~(42)  &
$ \Delta M_{4b,P_2}$             & -0.146~343~(35)  
\end{renomtable}
\end{ruledtabular}
\end{table}
\renewcommand{\arraystretch}{1}

Substituting the numerical results of the integrals listed in Tables~\ref{table:setII(c)_eee}
and \ref{table:setII(c)residual_eee}
in (\ref{eq:residualSetII(c)}),
we obtain
\begin{equation}
A_1^{(10)}[\text{Set~II(c)}^{(eee)}] = -0.116~489~(32).
\label{eq:setII(c)result}
\end{equation} 
We also calculated the contribution of Set~II(c) using the 
exact spectral function of $\Pi^{(4,2)}$ \cite{Hoang:1995ex}.
In this case  
$\Delta M_{4a ,P_{4 (P_2)} }^{(eee)}$, 
$\Delta M_{4b ,P_{4 (P_2)} }^{(eee)}$, $ M_{2, P_{4(P_2)}}^{(eee)}$,
and $\Delta B_{2,P_{4 (P_2)}}^{(eee)}$ 
in the right-hand side of Eq.~(\ref{eq:residualSetII(c)})  
can be directly evaluated using the exact spectral function.
The results are listed in the last two lines of Table~\ref{table:setII(c)_eee}
and in the fourth lines of Table~\ref{table:setII(c)residual_eee}.
The value obtained using the numbers in Table~\ref{table:setII(c)_eee} 
and Table~\ref{table:setII(c)residual_eee} is
\begin{equation}
A_1^{(10)}[\text{Set~II(c)}^{(eee)}: \text{spectral function} ] = -0.116~447~(34)~,  
\label{eq:setII(c)spectral}
\end{equation} 
which is in good agreement with (\ref{eq:setII(c)result}).
This shows that  {\sc gencodevp}{\it N} works correctly for $N=4$.


\subsection{Numerical results: $(eme)$, $(eee)$, etc.}
\label{subsec:eemresults}


Diagrams of Set~II(c) contain two closed lepton loops, one within the other.
We obtain mass-dependent contributions to the electron $g\!-\!2$ when
one or both loops consist of muon or tau-lepton.
The largest mass-dependent contributions come from 
the integral (\ref{eq:residualSetII(c)}) with superscripts $(eme)$ and 
then with $(eem)$. 
Results of numerical integration are listed in Table~\ref{table:setII(c)_eme_eem}.
The value obtained using the numbers in Tables~\ref{table:setII(c)_eme_eem} and \ref{table:setII(c)residual_eme_eem} are 
\begin{equation}
A_2^{(10)}[\text{Set~II(c)}^{(eme)}] =-0.260~86~(45) \times 10^{-3} ,
\label{eq:setII(c)result_eme}
\end{equation} 
and
\begin{equation}
A_2^{(10)}[\text{Set~II(c)}^{(eem)}] =-0.102~63~(14) \times 10^{-3} .
\label{eq:setII(c)result_eem}
\end{equation} 
Other mass-dependent terms of Set~II(c)
are listed in Table~\ref{table:setII(c)mass-dep-a_e}.

\renewcommand{\arraystretch}{0.80}
\begin{table}
\begin{ruledtabular}
\caption{ Contributions of diagrams
of Set~II(c) containing one electron loop and one muon loop to $a_e$.
The superscript $(eme)$ denotes a diagram in which the outer loop
is muon loop.
The superscript $(eem)$ denotes a diagram in which the inner loop
is muon loop.
The multiplicity of the
diagram $n_F$ is included in the numerical results. 
All integrals are evaluated in double precision. 
\label{table:setII(c)_eme_eem}
}
\begin{resulttable}
$\Delta M_{4a,P_{4a(P_2)}}^{(eme)}$&6 &  0.000~017~57~(22) &$1 \times 10^7$ & 50 \\
$\Delta M_{4a,P_{4b(P_2)}}^{(eme)}$&12&  0.000~013~37~(31) &$1 \times 10^7$ & 50 \\
$\Delta M_{4b,P_{4a(P_2)}}^{(eme)}$&6 & -0.000~172~92~(13) &$1 \times 10^7$ & 50 \\
$\Delta M_{4b,P_{4b(P_2)}}^{(eme)}$&12& -0.000~134~62~(18) &$1 \times 10^7$ & 50 \\
  & & & & \\
$\Delta M_{4a,P_{4a(P_2)}}^{(eem)}$&6 &  0.000~014~66~(11) &$1 \times 10^7$ & 50 \\
$\Delta M_{4a,P_{4b(P_2)}}^{(eem)}$&12&  0.000~001~73~( 3) &$1 \times 10^7$ & 50 \\
$\Delta M_{4b,P_{4a(P_2)}}^{(eem)}$&6 & -0.000~094~42~( 8) &$1 \times 10^7$ & 50 \\
$\Delta M_{4b,P_{4b(P_2)}}^{(eem)}$&12& -0.000~001~11~( 2) &$1 \times 10^7$ & 50  \\
\end{resulttable}
\end{ruledtabular}
\end{table}
\renewcommand{\arraystretch}{1}


\renewcommand{\arraystretch}{0.70}
\begin{table}
\begin{ruledtabular}
\caption{ Finite renormalization constants 
needed for the mass-dependent terms $(e,m,e)$  and $(e,e,m)$ of Set~II(c).
The finite renormalization constants needed for this term
but not listed here can be found in Table 
\ref{table:setII(c)residual_eee}.
All integrals are evaluated in double precision.
The multiplicity of the integral  
is incorporated in the numerical value. 
\label{table:setII(c)residual_eme_eem}
}
\begin{renomtable}
 $\Delta B_{2,P_{4a(P_2)}}^{(eme)}$     & 0.773~326~(53)   \times 10^{-4}   &
 $ M_{2,P_{4a(P_2)}}^{(eme)}$           & 0.044~986~(3) \times 10^{-4}  \\
 $\Delta B_{2,P_{4b(P_2)}}^{(eme)}$     & 0.603~167~(106)   \times 10^{-4}  &
 $ M_{2,P_{4b(P_2)}}^{(eme)}$           & 0.033~465~(6) \times 10^{-4} \\
                                  & & & \\                               
 $\Delta B_{2,P_{4a(P_2)}}^{(eem)}$     & 0.414~245~(42)\times 10^{-4}   &
 $ M_{2,P_{4a(P_2)}}^{(eem)}$           & 0.050~072~(5) \times 10^{-4}  \\
 $\Delta B_{2,P_{4b(P_2)}}^{(eem)}$     & 0.005~000~(9) \times 10^{-4}  &
 $ M_{2,P_{4b(P_2)}}^{(eem)}$           & 0.000~552~(1) \times 10^{-4} \\
                                  & & & \\                               
 $\Delta M_{4a,P_2}^{(em)}$             & 0.020~90~(21)    \times 10^{-4} &
 $\Delta M_{4b,P_2}^{(em)}$             &-0.209~84~(12)    \times 10^{-4}  \\ 
 $\Delta B_{2,P_2}^{(em)}$              & 0.094~050~(3) \times 10^{-4}  &
 $ M_{2,P_2}^{(em)}$                    & 0.005~197~62~(21) \times 10^{-4} \\  
 $\Delta B_{2,P_2}^{(me)}$              & 1.885~69~(24)               &
                                        &                              
\end{renomtable}
\end{ruledtabular}
\end{table}
\renewcommand{\arraystretch}{1}

\renewcommand{\arraystretch}{0.80}
\begin{table}
\begin{ruledtabular}
\caption{ Mass-dependent contributions of diagrams of Set~II(c)
to the electron $g\!-\!2$.
All integrals are evaluated in double precision.
\label{table:setII(c)mass-dep-a_e}
}
\begin{tabular}{clcl}
\multicolumn{1}{c}{$(e,l_2, l_3)$}
 & \multicolumn{1}{c}{$A_2^{(10)(el_2l_3)}$} &
 \multicolumn{1}{c}{$(e,l_2, l_3)$}
 & \multicolumn{1}{c}{$A_3^{(10)(el_2l_3)}$} \\ 
\hline 
$(e,m,m)$ & $-0.167~65~(28)\times 10^{-4}$  
         & $(e,m,t)$ & $-0.410~01~(81)\times 10^{-6}$ \\
$(e,t,e)$ & $-0.287~97~(58) \times 10^{-5}$ 
         & $(e,t,m)$ & $-0.784~4~(14)\times 10^{-6}$ \\
$(e,e,t)$ & $-0.988~9~(20) \times 10^{-6}$  & & \\ 
$(e,t,t)$ & $-0.988~4~(21) \times 10^{-7}$  & &                           
\end{tabular}
\end{ruledtabular}
\end{table}
\renewcommand{\arraystretch}{1}

\subsection{Muon $g\!-\!2$: $(mee)$}
\label{subsec:meeresults}

The leading contribution to the muon $g\!-\!2$ comes from the 
case where both loops consist of electrons, namely, the $(mee)$ case,
where $m$ stands for the muon.
The value obtained using the numbers in Tables~\ref{table:setII(c)_mee} 
and \ref{table:setII(c)residual_mee_mme_mem} is
\begin{equation}
A_2^{(10)}[\text{Set~II(c)}^{(mee)}] = -3.888~27~(90).
\label{eq:setII(c)result_mee}
\end{equation} 
We checked this result  using the exact spectral function:
\begin{equation}
A_2^{(10)}[\text{Set~II(c)}^{(mee)}:\text{spectral function}] = -3.887~65~(92).
\label{eq:setII(c)result_mee_spf}
\end{equation}

\subsection{Muon $g\!-\!2$: $(mme)$, $(mem)$ and others}
\label{subsec:mme_and_mem_results}

The next-to-leading order contribution arises when the inner and outer loops
consist of electron and muon, respectively. We found
\begin{equation}
A_2^{(10)}[\text{Set~II(c)}^{(mme)}] = -1.345~98~(36).
\label{eq:setII(c)result_mme}
\end{equation}
When the inner and outer loops consists of muon and electron, respectively,
the contribution is found to be smaller:
\begin{equation}
A_2^{(10)}[\text{Set~II(c)}^{(mem)}] = -0.151~50~(15).
\label{eq:setII(c)result_mem}
\end{equation}

The contributions involving tau-lepton loops are summarized in
Table~\ref{table:setII(c)mass-dep-a_mu}.

\renewcommand{\arraystretch}{0.80}
\begin{table}
\begin{ruledtabular}
\caption{ Contributions to the muon $g\!-\!2$ from Set~II(c) diagrams 
involving closed electron and/or muon loops.
The multiplicity of the
diagram $n_F$ is included in the numerical results. 
All integrals are evaluated in double precision. 
\\
\label{table:setII(c)_mee}
}
\begin{resulttable}
$\Delta M_{4a,P_{4a(P_2)}}^{(mee)}$ &6 &  0.684~47~(37)&$1 \times 10^8,~1\times10^9$ &50,~40 \\
$\Delta M_{4a,P_{4b(P_2)}}^{(mee)}$ &12&  2.071~36~(55)&$1 \times 10^8,~1\times10^9$ &50,~50 \\
$\Delta M_{4b,P_{4a(P_2)}}^{(mee)}$ &6 &  0.025~50~(33)&$1 \times 10^8,~1\times10^9$ &50,~40 \\
$\Delta M_{4b,P_{4b(P_2)}}^{(mee)}$ &12& -4.320~77~(51)&$1 \times 10^8,~1\times10^9$ &50,~50 \\
                &  &               &                             & \\ 
$\Delta M_{4a,P_{4a(P_2)}}^{(mme)}$ &6 &  0.302~14~(11)&$1 \times 10^8,~1\times10^9$ &50,~10 \\
$\Delta M_{4a,P_{4b(P_2)}}^{(mme)}$ &12&  0.251~36~(19)&$1 \times 10^8,~1\times10^9$ &50,~10 \\
$\Delta M_{4b,P_{4a(P_2)}}^{(mme)}$ &6 & -0.957~39~( 9)&$1 \times 10^8,~1\times10^9$ &50,~10 \\
$\Delta M_{4b,P_{4b(P_2)}}^{(mme)}$ &12& -0.931~56~(15)&$1 \times 10^8,~1\times10^9$ &50,~10 \\
                 &  &               &                             & \\
$\Delta M_{4a,P_{4a(P_2)}}^{(mem)}$ &6 &  0.049~57~(10)&$1 \times 10^8$              &50 \\
$\Delta M_{4a,P_{4b(P_2)}}^{(mem)}$ &12&  0.001~64~( 5)&$1 \times 10^8$              &50 \\
$\Delta M_{4b,P_{4a(P_2)}}^{(mem)}$ &6 & -0.141~38~( 8)&$1 \times 10^8$              &50 \\
$\Delta M_{4b,P_{4b(P_2)}}^{(mem)}$ &12& -0.010~38~( 4)&$1 \times 10^8$              &50 \\
\end{resulttable}
\end{ruledtabular}
\end{table}
\renewcommand{\arraystretch}{1}


\renewcommand{\arraystretch}{0.70}
\begin{table}
\begin{ruledtabular}
\caption{ Finite renormalization terms of Set~II(c) for 
the muon anomaly $a_{\mu}$.
All integrals are evaluated in double precision.
The multiplicity of the integral  
is incorporated in the numerical value. 
\\
\label{table:setII(c)residual_mee_mme_mem}
}
\begin{renomtable}

 $\Delta B_{2,P_{4a(P_2)}}^{(mee)}$     & 0.655~71~(11)   &
 $ M_{2,P_{4a(P_2)}}^{(mee)}$           & 0.597~44~1~(48)  \\
 $\Delta B_{2,P_{4b(P_2)}}^{(mee)}$     & 2.279~41~(17)   &
 $ M_{2,P_{4b(P_2)}}^{(mee)}$           & 0.982~06~6~(70)  \\
 $\Delta B_{2,P_{4(P_2)}}^{(mee)}$      & 2.695~12~(64)   &
 $ M_{2,P_{4(P_2)}}^{(mee)}$            & 1.440~46~(28)  \\
         &   &   &         \\
 $\Delta B_{2,P_{4a(P_2)}}^{(mme)}$     & 0.417~691~(10)   &
 $ M_{2,P_{4a(P_2)}}^{(mme)}$           & 0.121~908~( 3)  \\
 $\Delta B_{2,P_{4b(P_2)}}^{(mme)}$     & 0.404~336~(22)   &
 $ M_{2,P_{4b(P_2)}}^{(mme)}$           & 0.099~237~( 5)  \\        
         &   &   &         \\
 $\Delta B_{2,P_{4a(P_2)}}^{(mem)}$     & 0.065~066~( 6)   &
 $ M_{2,P_{4a(P_2)}}^{(mem)}$           & 0.021~016~( 2)  \\
 $\Delta B_{2,P_{4b(P_2)}}^{(mem)}$     & 0.004~533~( 3)   &
 $ M_{2,P_{4b(P_2)}}^{(mem)}$           & 0.000~586~( 1)  \\ 
          &   &   &         \\
 $\Delta M_{4a,P_2}^{(me)}$     &  1.725~62~(49)  &
 $\Delta M_{4b,P_2}^{(me)}$     & -2.354~33~(45) \\  
 $\Delta B_{2,P_2}^{(me)}$      &  1.885~732~(16) &
 $ M_{2,P_2}^{(me)}$            &  1.094~258~282~7~(98) \\ 
\end{renomtable}
\end{ruledtabular}
\end{table}
\renewcommand{\arraystretch}{1}
\renewcommand{\arraystretch}{0.80}
\begin{table}
\begin{ruledtabular}
\caption{ Contributions to the muon $g\!-\!2$ from Set~II(c) diagrams involving tau-lepton loops.
All integrals are evaluated in double precision.
\label{table:setII(c)mass-dep-a_mu}
}
\begin{tabular}{clcl}
\multicolumn{1}{c}{$(m,l_2,l_3)$}
 & \multicolumn{1}{c}{$A_2^{(10)(m l_2 l_3)}$} &
 \multicolumn{1}{c}{$(m,l_2,l_3)$}
 & \multicolumn{1}{c}{$A_3^{(10)(m l_2 l_3)}$} \\ 
\hline 
$(m,m,t)$ & $-0.004~325~0~(49) $ & $(m,e,t)$ & $-0.004~734~1~(55)$ \\
$(m,t,m)$ & $-0.010~519~(13)   $ & $(m,t,e)$ & $-0.036~066~(51)$ \\
$(m,t,t)$ & $-0.001~504~1~(19) $ & & \\ 
\end{tabular}
\end{ruledtabular}
\end{table}
\renewcommand{\arraystretch}{1}


\section{Set~II(d)}
\label{sec:set2d}

Following the same consideration leading to Eq.~(\ref{eq:residualSetII(c)}) of Sec.~\ref{sec:set2c} we obtain
\begin{equation}
A_1^{(10)}[\text{Set~II(d)}^{(l_1l_2)}] =
         \sum_{i=a}^b \Delta M_{4i,P_6}^{(l_1l_2)}
              - \Delta B_2  M_{2,P_6}^{(l_1l_2)}
              - \Delta B_{2, P_6}^{(l_1l_2)}  M_2 ,
\label{eq:setII(d)residual}
\end{equation}
where  $l_2$ designates the loop lepton.
The vacuum-polarization function $P_6$ is fully renormalized
whose divergence structure can be readily found by the $K$-operation.
The renormalization formula for  $P_6$ takes different forms
depending on whether one follows the original $K$-operation 
prescription\cite{Kinoshita:1990}
or the $K$-operation plus $R$-subtraction method \cite{Aoyama:2007bs}.
In the first approach
the UV-finite part of fourth-order mass-renormalization term
is not subtracted when $P_{6C}$ and $P_{6D}$ are constructed.
In the second approach we 
subtract the mass-renormalization term completely,
including the finite part $\Delta \delta m_4$, which leads to
\begin{align}
\Delta M_{4i,P_6}^{(l_1l_2)}& = 
        \sum_{\beta=A}^{H} \Delta M_{4i,P_{6\beta}}^{(l_1l_2)} 
\nonumber \\
                  & - 4 \Delta B_2  \Delta M_{4i, P_4}^{(l_1l_2)} 
                    - 3 (\Delta B_2)^2 \Delta M_{4i,P_2}^{(l_1l_2)} 
        - 2\Delta L\!B_4 \Delta M_{4i,P_2}^{(l_1l_2)}, ~~~ \text{for}~i=a,b~,
\nonumber \\
M_{2,P_6}^{(l_1l_2)} & = 
       \sum_{\beta=A}^{H}  M_{2,P_{6\beta}}^{(l_1l_2)} 
\nonumber \\
                  & - 4 \Delta B_2  M_{2, P_4}^{(l_1l_2)} 
                    - 3 (\Delta B_2)^2  M_{2,P_2}^{(l_1l_2)}
                    - 2\Delta L\!B_4  M_{2,P_2}^{(l_1l_2)},
\nonumber \\
\Delta B_{2,P_6}^{(l_1l_2)} & = \sum_{\beta=A}^{H} \Delta B_{2,P_{6\beta}}^{(l_1l_2)} 
\nonumber \\
                  & - 4 \Delta B_2 \Delta B_{2, P_4}^{(l_1l_2)} 
                    - 3 (\Delta B_2)^2 \Delta B_{2,P_2}^{(l_1l_2)} 
                    - 2\Delta L\!B_4 \Delta B_{2,P_2}^{(l_1l_2)}.
\label{eq:M4P6}
\end{align}
The quantities in the right-hand-side of (\ref{eq:M4P6}), 
$\Delta M_{4i,P_{6\beta}}$, 
$\Delta B_{2,P_{6\beta}}$, and $ M_{2,P_{6\beta}}$ are 
defined by the $K$-operation and $R$-subtraction.
$\Delta L\!B_4$ is the sum of the finite parts of 
the fourth-order vertex-renormalization
constant $\Delta L_4$ and wave-function 
renormalization constant $\Delta B_4$. See Refs.~\cite{Kinoshita:1990,Aoyama:2007dv,Aoyama:2007mn} 
for the exact definition.  
Note that terms like $\Delta M_{4i,P_{6\beta}}$ include the multiplicity
$n_F$ of Feynman diagrams that contribute to them.

%

\subsection{Numerical results: $(ee)$ case}
\label{subsec:eeresults}

Preliminary calculations  of the $(ee)$ case based on Methods (a) and (b)
described in Sec.~\ref{subsec:eeeresults}
are consistent with each other
within the uncertainty estimated by VEGAS.
Therefore we list only the results of method (b) in Table~\ref{table:setII(d)_ee}.
From this table and Table~\ref{table:setII(d)residual_ee_em} we obtain
\begin{equation}
A_1^{(10)}[\text{Set~II(d)}^{(ee)}] = -0.243~00~(29)~.
\label{eq:setII(d)result}
\end{equation}
%
\renewcommand{\arraystretch}{0.80}
\begin{table}
\begin{ruledtabular}
\caption{ Contributions of diagrams
of Set~II(d), $(ee)$ case.
$n_F$ is the number of Feynman diagrams represented by the integral.
The fourth-order mass-renormalization is completed by $R$-subtraction 
within the
numerical programs of $\Delta M_{4a,P_{6C}}$, $\Delta M_{4a,P_{6D}}$,
$\Delta M_{4b,P_{6C}}$, and $\Delta M_{4b,P_{6D}}$.
All integrals are evaluated in double precision.
\label{table:setII(d)_ee}
}
\begin{resulttable}
$\Delta M_{4a,P_{6A}} $&12 &  0.112~990~(116) & $1\times10^8,~1\times10^9$ & 50,~100\\ 
$\Delta M_{4a,P_{6B}} $&6  &  0.072~919~( 72) & $1\times10^8,~1\times10^9$ & 50,~100\\ 
$\Delta M_{4a,P_{6C}} $&12 &  0.044~224~( 87) & $1\times10^8,~1\times10^9$ & 50,~100\\ 
$\Delta M_{4a,P_{6D}} $&12 & -0.088~822~( 78) & $1\times10^8,~1\times10^9$ & 50,~100\\ 
$\Delta M_{4a,P_{6E}} $&24 &  0.444~033~(113) & $1\times10^8,~1\times10^9$ & 50,~100\\ 
$\Delta M_{4a,P_{6F}} $&12 & -0.156~407~( 67) & $1\times10^8,~1\times10^9$ & 50,~100\\ 
$\Delta M_{4a,P_{6G}} $&6  &  0.094~162~( 54) & $1\times10^8,~1\times10^9$ & 50,~100\\ 
$\Delta M_{4a,P_{6H}} $&6  &  0.060~989~( 35) & $1\times10^8,~1\times10^9$ & 50,~100\\ 
$\Delta M_{4b,P_{6A}} $&12 & -0.398~926~( 66) & $1\times10^8,~1\times10^9$ & 50,~100\\ 
$\Delta M_{4b,P_{6B}} $&6  & -0.253~369~( 42) & $1\times10^8,~1\times10^9$ & 50,~100\\ 
$\Delta M_{4b,P_{6C}} $&12 & -0.141~941~( 51) & $1\times10^8,~1\times10^9$ & 50,~100\\ 
$\Delta M_{4b,P_{6D}} $&12 &  0.292~773~( 44) & $1\times10^8,~1\times10^9$ & 50,~100\\ 
$\Delta M_{4b,P_{6E}} $&24 & -1.395~971~( 66) & $1\times10^8,~1\times10^9$ & 50,~100\\ 
$\Delta M_{4b,P_{6F}} $&12 &  0.570~363~( 40) & $1\times10^8,~1\times10^9$ & 50,~100\\ 
$\Delta M_{4b,P_{6G}} $&6 & -0.232~467~( 32) & $1\times10^8,~1\times10^9$ & 50,~100\\ 
$\Delta M_{4b,P_{6H}} $&6 & -0.223~983~( 22) & $1\times10^8,~1\times10^9$ & 50,~100\\ 
\end{resulttable}
\end{ruledtabular}
\end{table}
\renewcommand{\arraystretch}{1}


\renewcommand{\arraystretch}{0.70}
\begin{table}
\begin{ruledtabular}
\caption{ Finite renormalization terms necessary for 
the cases $(ee)$ and $(em)$ of Set~II(d).
For simplicity the superscript $(ee)$ is omitted.
The values $M_{2,P_{6C}}$ and $M_{2,P_{6D}}$
are different from those in Table.~I of Ref.~\cite{Kinoshita:2005zr}.
The former is constructed with the $K$-operation and $R$-subtraction, while
the latter is with the $K$-operation only. 
All integrals are evaluated in double precision.
\label{table:setII(d)residual_ee_em}
}
\begin{renomtable}
$ \Delta M_{4a,P_4}$      &  0.131~298~(8)  &
$ \Delta M_{4b,P_4}$      & -0.420~295~(8)  \\
$ \Delta M_{4a,P_2}$      &  0.039~642~(42)  &
$ \Delta M_{4b,P_2}$      & -0.146~343~(35)  \\
$  M_{2,P_{6A}}$          &  0.044~446~7~(22) &
$ \Delta B_{2,P_{6A}}$    &  0.173~609~1~(96) \\
$  M_{2,P_{6B}}$          &  0.028~593~9~(14) &
$ \Delta B_{2,P_{6B}}$    &  0.110~466~1~(56) \\
$  M_{2,P_{6C}}$          &  0.017~717~3~(19) &
$ \Delta B_{2,P_{6C}}$    &  0.062~134~7~(74) \\
$  M_{2,P_{6D}}$          & -0.035~167~0~(16) &
$ \Delta B_{2,P_{6D}}$    & -0.127~657~6~(63) \\
$  M_{2,P_{6E}}$          &  0.179~333~2~(21) &
$ \Delta B_{2,P_{6E}}$    &  0.610~385~8~(84) \\
$  M_{2,P_{6F}}$          & -0.062~003~2~(12) &
$ \Delta B_{2,P_{6F}}$    & -0.247~658~3~(54) \\
$  M_{2,P_{6G}}$          &  0.038~879~0~(10) &
$ \Delta B_{2,P_{6G}}$    &  0.104~070~2~(44) \\
$  M_{2,P_{6H}}$          &  0.023~674~9~( 8) &
$ \Delta B_{2,P_{6H}}$    &  0.097~056~7~(29) \\
$  M_{2,P_{4}}$           &  0.052~870~652\cdots &
$ \Delta B_{2,P_{4}}$     &  0.183~666~8~(18) \\
$  M_{2,P_{2}}$           &  0.015~687~421\cdots &
$ \Delta B_{2,P_{2}}$     &  0.063~399~266\cdots \\
$ \Delta L\!B_4$          &  0.027~930~(27)  &
$ \Delta B_2$             &  0.75             \\
       $M_2$              &  0.5             & 
                          &                   \\
        &       &         &        \\
 $ \Delta M_{4a,P_4}^{(em)}$     &    0.075~96~(78)  \times 10^{-4}  &
 $ \Delta M_{4b,P_4}^{(em)}$     &   -0.757~35~(41)  \times 10^{-4}  \\
 $ \Delta M_{4a,P_2}^{(em)}$     &    0.020~90~(21)  \times 10^{-4} &
 $ \Delta M_{4b,P_2}^{(em)}$     &   -0.209~84~(12)  \times 10^{-4}  \\
 $  M_{2,P_{6A}}^{(em)}$         &    0.159~949 ~(32)\times 10^{-5} &
 $\Delta B_{2,P_{6A}}^{(em)}$    &    0.281~371 ~(61)\times 10^{-4}\\
 $  M_{2,P_{6B}}^{(em)}$         &    0.102~323 ~(21)\times 10^{-5} &
 $\Delta B_{2,P_{6B}}^{(em)}$    &    0.180~003 ~(40)\times 10^{-4}\\
 $  M_{2,P_{6C}}^{(em)}$         &    0.071~033 ~(29)\times 10^{-5} &
 $\Delta B_{2,P_{6C}}^{(em)}$    &    0.119~848 ~(52)\times 10^{-4}\\
 $  M_{2,P_{6D}}^{(em)}$         &   -0.131~368 ~(26)\times 10^{-5} &
 $\Delta B_{2,P_{6D}}^{(em)}$    &   -0.226~651 ~(46)\times 10^{-4}\\
 $  M_{2,P_{6E}}^{(em)}$         &    0.682~404 ~(32)\times 10^{-5} &
 $\Delta B_{2,P_{6E}}^{(em)}$    &    1.162~709 ~(59)\times 10^{-4}\\
 $  M_{2,P_{6F}}^{(em)}$         &   -0.211~985 ~(19)\times 10^{-5} &
 $\Delta B_{2,P_{6F}}^{(em)}$    &   -0.379~345 ~(35)\times 10^{-4}\\
 $  M_{2,P_{6G}}^{(em)}$         &    0.174~650 ~(16)\times 10^{-5} &
 $\Delta B_{2,P_{6G}}^{(em)}$    &    0.279~935 ~(31)\times 10^{-4}\\
 $  M_{2,P_{6H}}^{(em)}$         &    0.082~748 ~(12)\times 10^{-5} &
 $\Delta B_{2,P_{6H}}^{(em)}$    &    0.147~672 ~(22)\times 10^{-4}\\
$  M_{2,P_{4}}^{(em)}$           &  0.197~298~(5) \times 10^{-5} &
$ \Delta B_{2,P_{4}}^{(em)}$     &  0.338~738~(12)\times 10^{-4} \\
$  M_{2,P_{2}}^{(em)}$           &  0.051~974~(1) \times 10^{-5} &
$ \Delta B_{2,P_{2}}^{(em)}$     &  0.094~050~(3) \times 10^{-4} \\
\end{renomtable}
\end{ruledtabular}
\end{table}
\renewcommand{\arraystretch}{1}

The Pad\'{e}-approximated vacuum-polarization function of the sixth-order
with a single fermion loop has been obtained in Ref.~\cite{Baikov:1995ui,Kinoshita:1998jg}.
This method gives both imaginary and real parts of the vacuum-polarization function.  We use here only
its imaginary part to calculate its effect on the anomaly.
Numerical results of integration are summarized in Table.~\ref{table:setII(d)pade}.
Substituting them into Eq.~(\ref{eq:setII(d)residual}), we obtained
\begin{equation}
  A_1^{(10)}[\text{Set~II(d)}^{(ee)}:\text {Pad\'{e}}] = -0.243~06~(45)~,
\label{eq:setII(d)pade}
\end{equation}
which is in good agreement with (\ref{eq:setII(d)result}).
This provides another support for the validity of {\sc gencodevp}{\it N}.


\subsection{Numerical results: mass-dependent terms $(em)$ and $(et)$}
\label{subsec:emresults}

The value of the mass-dependent term $(em)$
obtained using the numbers listed in Table~\ref{table:setII(d)_em} is 
\begin{equation}
A_2^{(10)}[\text{Set~II(d)}^{(em)}] =-0.981~7~(42) \times 10^{-4} .
\label{eq:setII(d)result_em}
\end{equation} 

As a check we evaluated the same quantity using the Pad\'{e}-approximated
vacuum-polarization function of sixth-order.
The results are listed 
in Table~\ref{table:setII(d)pade}.  From these values we obtain
\begin{equation}
A_2^{(10)}[\text{Set~II(d)}^{(em)}: \text{Pad\'e}] =-0.991~53~(61) \times 10^{-4} .
\label{eq:setII(d)result_em_pade}
\end{equation} 
We also evaluated the mass-dependent term $(et)$
in Pad\'{e} approximation:
\begin{equation}
A_2^{(10)}[\text{Set~II(d)}^{(et)}: \text{Pad\'e}] =-0.542~7~(18) \times 10^{-6} .
\label{eq:setII(d)result_et_Pade}
\end{equation} 
We have not evaluated this term directly.
But it will be of the same order as
Eq.~(\ref{eq:setII(d)result_et_Pade}) and thus negligible numerically.

\renewcommand{\arraystretch}{0.80}
\begin{table}
\begin{ruledtabular}
\caption{ Contributions to $a_e$ from diagrams of Set~II(d) containing a muon loop.
The superscript $(em)$ signifies that the diagrams contain muon loop
in the electron $g\!-\!2$.
$n_F$ is the number of Feynman diagrams represented by the integral.
All integrals are evaluated in double precision.
The fourth-order mass renormalization is completed within the
numerical programs of $\Delta M_{4a,P_{6C}}$, $\Delta M_{4a,P_{6D}}$,
$\Delta M_{4b,P_{6C}}$, and $\Delta M_{4b,P_{6D}}$.
\label{table:setII(d)_em}
}
\begin{resulttable}
$\Delta M_{4a,P_{6A}}^{(em)} $&12 &  0.000~006~24~( 13)& $ 1\times10^7,~1\times10^8$ & 50,~50 \\ 
$\Delta M_{4a,P_{6B}}^{(em)} $&6 &  0.000~004~94~(  9)& $ 1\times10^7,~1\times10^8$ & 50,~50 \\ 
$\Delta M_{4a,P_{6C}}^{(em)} $&12 &  0.000~003~79~( 11)& $ 1\times10^7,~1\times10^8$ & 50,~50 \\ 
$\Delta M_{4a,P_{6D}}^{(em)} $&12 &  0.000~005~23~( 10)& $ 1\times10^7,~1\times10^8$ & 50,~50 \\ 
$\Delta M_{4a,P_{6E}}^{(em)} $&24 &  0.000~026~34~( 15)& $ 1\times10^7,~1\times10^8$ & 50,~50 \\ 
$\Delta M_{4a,P_{6F}}^{(em)} $&12 &  0.000~008~38~(  8)& $ 1\times10^7,~1\times10^8$ & 50,~50 \\ 
$\Delta M_{4a,P_{6G}}^{(em)} $&6 &  0.000~007~54~(  6)& $ 1\times10^7,~1\times10^8$ & 50,~50 \\ 
$\Delta M_{4a,P_{6H}}^{(em)} $&6 &  0.000~003~34~(  5)& $ 1\times10^7,~1\times10^8$ & 50,~50 \\ 
$\Delta M_{4b,P_{6A}}^{(em)} $&12 &  0.000~063~78~(  4)& $ 1\times10^7,~1\times10^8$ & 50,~50 \\ 
$\Delta M_{4b,P_{6B}}^{(em)} $&6 &  0.000~040~16~(  2)& $ 1\times10^7,~1\times10^8$ & 50,~50 \\ 
$\Delta M_{4b,P_{6C}}^{(em)} $&12 &  0.000~027~79~(  3)& $ 1\times10^7,~1\times10^8$ & 50,~50 \\ 
$\Delta M_{4b,P_{6D}}^{(em)} $&12 &  0.000~051~62~(  3)& $ 1\times10^7,~1\times10^8$ & 50,~50 \\ 
$\Delta M_{4b,P_{6E}}^{(em)} $&24 &  0.000~260~81~(  4)& $ 1\times10^7,~1\times10^8$ & 50,~50 \\ 
$\Delta M_{4b,P_{6F}}^{(em)} $&12 &  0.000~085~61~(  2)& $ 1\times10^7,~1\times10^8$ & 50,~50 \\ 
$\Delta M_{4b,P_{6G}}^{(em)} $&6 &  0.000~063~72~(  2)& $ 1\times10^7,~1\times10^8$ & 50,~50 \\ 
$\Delta M_{4b,P_{6H}}^{(em)} $&6 &  0.000~033~95~(  1)& $ 1\times10^7,~1\times10^8$ & 50,~50 \\ 
\end{resulttable}
\end{ruledtabular}
\end{table}
\renewcommand{\arraystretch}{1}

\renewcommand{\arraystretch}{0.80}
\begin{table}
\begin{ruledtabular}
\caption{Contributions of diagrams
of Set~II(d) whose VP function is $P_6$.
Quantities on the right-hand-side of Eq.~(\ref{eq:M4P6})
are calculated from Tables~\ref{table:setII(d)_ee}, 
\ref{table:setII(d)residual_ee_em}, \ref{table:setII(d)_em}, 
\ref{table:setII(d)_me}, \ref{table:setII(d)_mt}, 
and \ref{table:setII(d)residual_me_mt}.
The same quantities are also calculated by the Pad\'{e} approximant method
and listed with the subscript Pad\'{e} below the corresponding integrals. 
$ M_{2,P6}^{(l_1l_2)}$ are actually the
anomaly contributions of the eighth-order diagrams of Group I(d). 
Their values for the $(ee)$ and $(me)$ cases
are consistent with Eq.~(29) of Ref.~\cite{Kinoshita:2005zr} and Eq.~(34) of
Ref.~\cite{Kinoshita:2005sm}, respectively. The $(em)$, $(et)$, and $(mt)$ cases are
newly evaluated in this paper.
All integrals using the Pad\'{e} approximant are evaluated 
in quadruple precision.  
\label{table:setII(d)pade}
}
\begin{renomtable}
$\Delta M_{4a,P_6}^{(ee)}$              & 0.119~08~(25)   &
$\Delta M_{4b,P_6}^{(ee)}$              &-0.264~51~(41)   \\ 
$\Delta M_{4a,P_6~\text{Pad\'e}}^{(ee)}$& 0.118~95~(35)   &
$\Delta M_{4b,P_6~\text{Pad\'e}}^{(ee)}$&-0.264~43~(27)   \\ 

$\Delta B_{2,P_6}^{(ee)} $              & 0.120~879~(20)  & 
$ M_{2,P_6}^{(ee)} $              & 0.049~514~( 5)  \\ 
$\Delta B_{2,P_6~\text{Pad\'e}}^{(ee)} $& 0.120~862~(39)  & 
$ M_{2,P_6~\text{Pad\'e}}^{(ee)} $& 0.049~520~( 4)  \\ 
 & & & \\
$\Delta M_{4a,P_6}^{(em)}$              & 0.091~5 ~(39)  \times 10^{-4} &
$\Delta M_{4b,P_6}^{(em)}$              &-0.862~0 ~(16)  \times 10^{-4} \\ 
$\Delta M_{4a,P_6~\text{Pad\'e}}^{(em)}$& 0.091~96~(53)  \times 10^{-4} &
$\Delta M_{4b,P_6~\text{Pad\'e}}^{(em)}$&-0.872~25~(31)  \times 10^{-4} \\ 
$\Delta B_{2,P_6}^{(em)} $              & 0.385~36~(13)  \times 10^{-4} & 
$ M_{2,P_6}^{(em)} $              & 0.024~725~(7)  \times 10^{-4} \\ 
$\Delta B_{2,P_6~\text{Pad\'e}}^{(em)} $& 0.385~367~(72) \times 10^{-4} & 
$ M_{2,P_6~\text{Pad\'e}}^{(em)} $& 0.024~727~(4)  \times 10^{-4} \\ 
 & & & \\
$\Delta M_{4a,P_6~\text{Pad\'e}}^{(et)}$& 0.039~89~(154) \times 10^{-6} &
$\Delta M_{4b,P_6~\text{Pad\'e}}^{(et)}$&-0.470~90~(74)  \times 10^{-6} \\
$\Delta B_{2,P_6~\text{Pad\'e}}^{(et)} $& 0.210~278~(39) \times 10^{-6} &
$ M_{2,P_6~\text{Pad\'e}}^{(et)} $& 0.008~744~(1)  \times 10^{-6} \\ 
  & & & \\
$\Delta M_{4a,P_6}^{(me)}$              &-0.469~4~(50)   &
$\Delta M_{4b,P_6}^{(me)}$              & 0.596~9~(42)   \\
$\Delta M_{4a,P_6~\text{Pad\'e}}^{(me)}$&-0.459~1~(55)   &
$\Delta M_{4b,P_6~\text{Pad\'e}}^{(me)}$& 0.593~5~(51)   \\
$\Delta B_{2,P_6}^{(me)} $              &-0.394~72~(82)  & 
$ M_{2,P_6}^{(me)} $              &-0.229~82~(36)  \\ 
$\Delta B_{2,P_6~\text{Pad\'e}}^{(me)} $&-0.395~25~(78)  & 
$ M_{2,P_6~\text{Pad\'e}}^{(me)} $&-0.230~23~(32)  \\ 
 & & & \\
$\Delta M_{4a,P_6}^{(mt)}$              & 0.001~066~(15)   &
$\Delta M_{4b,P_6}^{(mt)}$              &-0.006~953~(10)    \\ 
$\Delta M_{4a,P_6~\text{Pad\'e}}^{(mt)}$& 0.001~065~(12)   &
$\Delta M_{4b,P_6~\text{Pad\'e}}^{(mt)}$&-0.006~953~(8)    \\ 
$\Delta B_{2,P_6}^{(mt)} $              & 0.003~020~9~(10)  & 
$ M_{2,P_6}^{(mt)} $              & 0.000~367~7~(2)  \\  
$\Delta B_{2,P_6~\text{Pad\'e}}^{(mt)} $& 0.003~021~4~(6)  & 
$ M_{2,P_6~\text{Pad\'e}}^{(mt)} $& 0.000~367~7~(1)    
\end{renomtable}
\end{ruledtabular}
\end{table}
\renewcommand{\arraystretch}{1}
%
%
  


\subsection{Muon $g\!-\!2$: $(me)$}
\label{subsec:meresults}


The leading contribution to the muon $g\!-\!2$ comes from the case $(me)$.
The value obtained using the numbers in Tables~\ref{table:setII(d)_me}, 
 \ref{table:setII(d)residual_ee_em}, and
\ref{table:setII(d)residual_me_mt} is
\begin{equation}
A_2^{(10)}[\text{Set~II(d)}^{(me)}] = 0.497~2 ~(65).
\label{eq:setII(d)result_me}
\end{equation} 
This is in fair agreement with the value obtained using the Pad\'{e}
approximant
\begin{equation}
A_2^{(10)}[\text{Set~II(d)}^{(me)}: {\text{Pad\'{e}}}] = 0.504~8 ~(75).
\label{eq:setII(d)result_me_pade}
\end{equation} 

A crude evaluation of the contribution of tau-lepton loop gives
\begin{equation}
A_2^{(10)}[\text{Set~II(d)}^{(mt)}] = -0.007~673~ (18),
\label{eq:setII(d)result_mt}
\end{equation} 
while the same contribution obtained using the Pad\'{e} approximant is
\begin{equation}
A_2^{(10)}[\text{Set~II(d)}^{(mt)}: {\text{Pad\'{e}}}] = -0.007~674~ (15),
\label{eq:setII(d)result_mt_pade}
\end{equation}

\renewcommand{\arraystretch}{0.80}
\begin{table}
\begin{ruledtabular}
\caption{ Leading contributions of diagrams
of Set~II(d) of Fig.~\ref{fig:X2cd} to the muon $g\!-\!2$.
$n_F$ is the number of Feynman diagrams represented by the integral.
All integrals are evaluated in double precision.
The fourth-order mass-renormalization is completed within the
numerical programs of $\Delta M_{4a,P_{6C}}$, $\Delta M_{4a,P_{6D}}$,
$\Delta M_{4b,P_{6C}}$, and $\Delta M_{4b,P_{6D}}$.
Last two columns list initial sampling and its iteration, followed by increased
sampling and its iteration.
\label{table:setII(d)_me}
}
\begin{resulttable}
$\Delta M_{4a,P_{6A}}^{(me)}$ & 12 &   10.693~5~(20)   & $1\times 10^8,~1\times 10^9,~1\times10^{10}$&50,~~100,~~185\\
$\Delta M_{4a,P_{6B}}^{(me)}$ & 6 &    5.642~1~(12)   & $1\times 10^8,~1\times 10^9,~1\times10^{10}$&50,~~100,~~170\\
$\Delta M_{4a,P_{6C}}^{(me)}$ & 12 &    4.305~0~(12)   & $1\times 10^8,~1\times 10^9,~1\times10^{10}$&50,~~100,~~170\\
$\Delta M_{4a,P_{6D}}^{(me)}$ & 12 &   -6.097~7~(12)   & $1\times 10^8,~1\times 10^9,~1\times10^{10}$&50,~~100,~~170\\
$\Delta M_{4a,P_{6E}}^{(me)}$ & 24 &   -4.365~9~(18)   & $1\times 10^8,~1\times 10^9,~1\times10^{10}$&50,~~100,~~170\\
$\Delta M_{4a,P_{6F}}^{(me)}$ & 12 &   -6.145~99~( 94) & $1\times 10^8,~1\times 10^9,~1\times10^{10}$&50,~~100,~~170\\
$\Delta M_{4a,P_{6G}}^{(me)}$ & 6 &   -0.397~7~(14)   & $1\times 10^8,~1\times 10^9,~1\times10^{10}$&50,~~100,~~170\\
$\Delta M_{4a,P_{6H}}^{(me)}$ & 6 &    5.048~76~( 52) & $1\times 10^8,~1\times 10^9,~1\times10^{10}$&50,~~100,~~100\\
$\Delta M_{4b,P_{6A}}^{(me)}$ & 12 &  -18.959~1~(16)   & $1\times 10^8,~1\times 10^9,~1\times10^{10}$&50,~~100,~~170\\
$\Delta M_{4b,P_{6B}}^{(me)}$ & 6 &   -9.815~33~( 96) & $1\times 10^8,~1\times 10^9,~1\times10^{10}$&50,~~100,~~150\\
$\Delta M_{4b,P_{6C}}^{(me)}$ & 12 &   -7.998~86~( 93) & $1\times 10^8,~1\times 10^9,~1\times10^{10}$&50,~~100,~~150\\
$\Delta M_{4b,P_{6D}}^{(me)}$ & 12 &   11.248~41~( 81) & $1\times 10^8,~1\times 10^9,~1\times10^{10}$&50,~~100,~~150\\
$\Delta M_{4b,P_{6E}}^{(me)}$ & 24 &   15.905~0~(12)   & $1\times 10^8,~1\times 10^9,~1\times10^{10}$&50,~~100,~~170\\
$\Delta M_{4b,P_{6F}}^{(me)}$ & 12 &    7.177~72~( 74) & $1\times 10^8,~1\times 10^9,~1\times10^{10}$&50,~~100,~~150\\
$\Delta M_{4b,P_{6G}}^{(me)}$ & 6 &   -0.501~22~( 94) & $1\times 10^8,~1\times 10^9,~1\times10^{10}$&50,~~100,~~160\\
$\Delta M_{4b,P_{6H}}^{(me)}$ & 6 &   -8.025~32~( 40) & $1\times 10^8,~1\times 10^9,~1\times10^{10}$&50,~~100,~~100\\
\end{resulttable}
\end{ruledtabular}
\end{table}
\renewcommand{\arraystretch}{1}


\renewcommand{\arraystretch}{0.80}
\begin{table}
\begin{ruledtabular}
\caption{ Contribution  from diagrams
of Set~II(d) $(m,t)$ of Fig.~\ref{fig:X2cd} to the muon $g\!-\!2$.
$n_F$ is the number of Feynman diagrams represented by the integral.
All integrals are evaluated in double precision.
Last two columns list initial sampling and its iteration, followed by increased
sampling and its iteration.
\label{table:setII(d)_mt}
}
\begin{resulttable}
$\Delta M_{4a,P_{6A}}^{(mt)} $&12 &     0.001~746~1~(22)   & $ 1\times10^8,~1\times10^9$& 50,~100\\ 
$\Delta M_{4a,P_{6B}}^{(mt)} $&6 &     0.000~476~4~(14)   & $ 1\times10^8,~1\times10^9$& 50,~100\\ 
$\Delta M_{4a,P_{6C}}^{(mt)} $&12 &     0.000~323~9~(18)   & $ 1\times10^8,~1\times10^9$& 50,~100\\ 
$\Delta M_{4a,P_{6D}}^{(mt)} $&12 &    -0.001~605~3~(17)   & $ 1\times10^8,~1\times10^9$& 50,~100\\ 
$\Delta M_{4a,P_{6E}}^{(mt)} $&24 &     0.003~112~4~(24)   & $ 1\times10^8,~1\times10^9$& 50,~100\\ 
$\Delta M_{4a,P_{6F}}^{(mt)} $&12 &    -0.002~000~0~(13)   & $ 1\times10^8,~1\times10^9$& 50,~100\\ 
$\Delta M_{4a,P_{6G}}^{(mt)} $&6 &     0.001~766~5~(12)   & $ 1\times10^8,~1\times10^9$& 50,~100\\ 
$\Delta M_{4a,P_{6H}}^{(mt)} $&6 &     0.000~390~55~( 72) & $ 1\times10^8,~1\times10^9$& 50,~100\\ 
$\Delta M_{4b,P_{6A}}^{(mt)} $&12 &    -0.006~549~50~( 87) & $ 1\times10^8,~1\times10^9$& 50,~100\\ 
$\Delta M_{4b,P_{6B}}^{(mt)} $&6 &    -0.004~550~21~( 57) & $ 1\times10^8,~1\times10^9$& 50,~100\\ 
$\Delta M_{4b,P_{6C}}^{(mt)} $&12 &    -0.002~299~84~( 74) & $ 1\times10^8,~1\times10^9$& 50,~100\\ 
$\Delta M_{4b,P_{6D}}^{(mt)} $&12 &     0.004~413~20~( 67) & $ 1\times10^8,~1\times10^9$& 50,~100\\ 
$\Delta M_{4b,P_{6E}}^{(mt)} $&24 &    -0.022~437~50~( 93) & $ 1\times10^8,~1\times10^9$& 50,~100\\ 
$\Delta M_{4b,P_{6F}}^{(mt)} $&12 &     0.008~566~26~( 53) & $ 1\times10^8,~1\times10^9$& 50,~100\\ 
$\Delta M_{4b,P_{6G}}^{(mt)} $&6 &    -0.005~164~07~( 46) & $ 1\times10^8,~1\times10^9$& 50,~100\\ 
$\Delta M_{4b,P_{6H}}^{(mt)} $&6 &    -0.003~940~06~( 31) & $ 1\times10^8,~1\times10^9$& 50,~100\\ 
\end{resulttable}
\end{ruledtabular}
\end{table}
\renewcommand{\arraystretch}{1}


\renewcommand{\arraystretch}{0.70}
\begin{table}
\begin{ruledtabular}
\caption{ Finite renormalization terms needed for 
Set~II(d) of the muon $g\!-\!2$.
Both $(me)$ and $(mt)$ cases are listed.
$ M_{2,P_{6\beta}}^{(me)}$ with $\beta = A,B,E,F,G,H$  are 
consistent with
those in  Ref.~\cite{Kinoshita:2005sm}.  $ M_{2, P_{6\beta}}$ with $\beta=C,D$ 
in this table incorporated  the $R$-subtraction \cite{Aoyama:2007bs} and differs
from those in Ref.~\cite{Kinoshita:2005sm}. 
\label{table:setII(d)residual_me_mt}
}
\begin{renomtable}
$ \Delta M_{4a,P_4}^{(me)}$     &    2.047~84~(23)    &
$ \Delta M_{4b,P_4}^{(me)}$     &   -2.486~60~(12)    \\
$ \Delta M_{4a,P_2}^{(me)}$     &    1.725~62~(49)    &
$ \Delta M_{4b,P_2}^{(me)}$     &   -2.354~33~(46)    \\
$  M_{2,P_{6A}}^{(me)}$         &  5.676~49~(22) &
$ \Delta B_{2,P_{6A}}^{(me)}$   & 11.500~82~(47) \\
$  M_{2,P_{6B}}^{(me)}$         &  3.058~13~(13) &
$ \Delta B_{2,P_{6B}}^{(me)}$   &  6.101~83~(27) \\
$  M_{2,P_{6C}}^{(me)}$         &  2.194~66~(12) &
$ \Delta B_{2,P_{6C}}^{(me)}$   &  4.616~02~(24) \\
$  M_{2,P_{6D}}^{(me)}$         & -3.224~25~(10) &
$ \Delta B_{2,P_{6D}}^{(me)}$   & -6.701~64~(22) \\
$  M_{2,P_{6E}}^{(me)}$         & -0.073~76~(17) &
$ \Delta B_{2,P_{6E}}^{(me)}$   & -4.076~72~(36) \\
$  M_{2,P_{6F}}^{(me)}$         & -4.064~09~( 9) &
$ \Delta B_{2,P_{6F}}^{(me)}$   & -6.535~49~(20) \\
$  M_{2,P_{6G}}^{(me)}$         & -0.246~97~(12) &
$ \Delta B_{2,P_{6G}}^{(me)}$   & -0.039~85~(28) \\
$  M_{2,P_{6H}}^{(me)}$         &  2.838~67~( 4) &
$ \Delta B_{2,P_{6H}}^{(me)}$   &  5.345~15~( 8) \\
$  M_{2,P_{4}}^{(me)}$          &  1.493~671~581~(8) &
$ \Delta B_{2,P_{4}}^{(me)}$    &  2.439~109~(53) \\
$  M_{2,P_{2}}^{(me)}$          &  1.094~258~282~7~(98) &
$ \Delta B_{2,P_{2}}^{(me)}$    &  1.885~733~(16) \\
         & & & \\
$ \Delta M_{4a,P_4}^{(mt)}$      &  0.000~903~9~(47)    &
$ \Delta M_{4b,P_4}^{(mt)}$      & -0.006~571~6~(31)    \\
$ \Delta M_{4a,P_2}^{(mt)}$      &  0.000~247~8~(15)     &
$ \Delta M_{4b,P_2}^{(mt)}$      & -0.001~888~8~(10)    \\
$  M_{2,P_{6A}}^{(mt)}$          &  0.000~239~71~( 1) &  
$ \Delta B_{2,P_{6A}}^{(mt)}$    &  0.002~436~62~(13) \\ 
$  M_{2,P_{6B}}^{(mt)}$          &  0.000~153~39~( 1) & 
$ \Delta B_{2,P_{6B}}^{(mt)}$    &  0.001~558~94~( 8) \\ 
$  M_{2,P_{6C}}^{(mt)}$          &  0.000~106~19~( 1) & 
$ \Delta B_{2,P_{6C}}^{(mt)}$    &  0.001~006~23~(11) \\ 
$  M_{2,P_{6D}}^{(mt)}$          & -0.000~196~75~( 1) & 
$ \Delta B_{2,P_{6D}}^{(mt)}$    & -0.001~934~79~( 9) \\ 
$  M_{2,P_{6E}}^{(mt)}$          &  0.001~021~35~( 1) & 
$ \Delta B_{2,P_{6E}}^{(mt)}$    &  0.009~826~56~(12) \\ 
$  M_{2,P_{6F}}^{(mt)}$          & -0.000~318~06~( 1) & 
$ \Delta B_{2,P_{6F}}^{(mt)}$    & -0.003~326~19~( 7) \\ 
$  M_{2,P_{6G}}^{(mt)}$          &  0.000~260~44~( 1) & 
$ \Delta B_{2,P_{6G}}^{(mt)}$    &  0.002~250~09~( 6) \\ 
$  M_{2,P_{6H}}^{(mt)}$          &  0.000~124~05~( 1) & 
$ \Delta B_{2,P_{6H}}^{(mt)}$    &  0.001~292~36~( 4) \\ 
$  M_{2,P_{4}}^{(mt)}$           &  0.000~295~508~(21) &
$ \Delta B_{2,P_{4}}^{(mt)}$     &  0.002~880~01~(31) \\
$  M_{2,P_{2}}^{(mt)}$           &  0.000~078~067~4~(31) &
$ \Delta B_{2,P_{2}}^{(mt)}$     &  0.000~831~107~(75) \\
\end{renomtable}
\end{ruledtabular}
\end{table}
\renewcommand{\arraystretch}{1}

%

\section{summary and discussion}
\label{sec:summary}

The total contribution to $a_e$ from the set II(c)
is the sum of Eqs.~(\ref{eq:setII(c)result}),
(\ref{eq:setII(c)result_eme}),
(\ref{eq:setII(c)result_eem}),
and 
other terms listed in Table~\ref{table:setII(c)mass-dep-a_e}:
\begin{equation}
a_e^{(10)}[\text{Set~II(c):all}] = -0.116~874~(43) \left( \frac{\alpha}{\pi} \right)^5 .
\label{eq:setII(c)all}
\end{equation} 
%
Contributions of tau-lepton loop listed in Table~\ref{table:setII(c)mass-dep-a_mu}
are less than the uncertainty of
Eq.~(\ref{eq:setII(c)all}).

The total contribution to $a_e$ from the Set~II(d)
is the sum of Eq.~(\ref{eq:setII(d)result})
and Eq.~(\ref{eq:setII(d)result_em}):
\begin{equation}
a_e^{(10)}[\text{Set~II(d):all}] = -0.243~10~(29) \left( \frac{\alpha}{\pi} \right)^5 .
\label{eq:setII(d)all}
\end{equation} 
The contribution of the tau-lepton loop is 
within the error bars of Eq.~(\ref{eq:setII(d)all})
and is completely negligible at present.

The total contribution of Set~II(c) to the muon $g\!-\!2$
involving electron, muon, and tau-lepton loops
is the sum of Eq.~(\ref{eq:setII(c)result}), 
Eq.~(\ref{eq:setII(c)result_mee}), Eq.~(\ref{eq:setII(c)result_mme}), 
and Eq.~(\ref{eq:setII(c)result_mem}), and values listed 
in Table.~\ref{table:setII(c)mass-dep-a_e}:
\begin{equation}
a_\mu^{(10)}[\text{Set~II(c):all}] = -5.559~4~(11) \left( \frac{\alpha}{\pi} \right)^5 .
\label{eq:setII(c)mu}
\end{equation} 

The total contribution of Set~II(d) to the muon $g\!-\!2$
involving electron, muon, and tau-lepton loops
is the sum of 
Eq.~(\ref{eq:setII(d)result}), Eq.~(\ref{eq:setII(d)result_me}), 
and Eq.~(\ref{eq:setII(d)result_mt}):
\begin{equation}
a_\mu^{(10)}[\text{Set~II(d):all}] =  0.246~5~(65) \left( \frac{\alpha}{\pi} \right)^5 .
\label{eq:setII(d)mu}
\end{equation} 

The sum of the electron-loop contribution to the muon $g\!-\!2$ from
the diagrams Set~II(c) and Set~II(d) is the sum of
Eqs.(\ref{eq:setII(c)result_mee}), (\ref{eq:setII(c)result_mme}),
(\ref{eq:setII(c)result_mem}), and (\ref{eq:setII(d)result_me}). We find
\begin{equation}
A_2^{(10)}(m_\mu/m_e)[\text{Set~II(c+d)}^{(me)}] =  -4.888~6~(65)  ,
\label{eq:setII(c+d)}
\end{equation} 
which 
is less than 1\% of the leading contribution from
the diagrams of Set VI(a)
that contain light-by-light-scattering subdiagrams and vacuum-polarization
subdiagrams\cite{Kinoshita:2005sm,Kataev:2006yh}. 
Hence, the new contribution does not alter the previous estimates:  
\begin{eqnarray}
&& A_2^{(10)}(m_\mu/m_e)[\text{estimate}:\text{Ref.\cite{Kinoshita:2005sm}}] 
               =  663~(20)~, 
\\
&& A_2^{(10)}(m_\mu/m_e)[\text{estimate}:\text{Ref.\cite{Kataev:2006yh}}] =  643~(20)~. 
\label{eq:A_2(mu/me_estimate)}
\end{eqnarray}


\begin{acknowledgments}
This work is supported in part by the JSPS Grant-in-Aid for
Scientific Research (C)19540322 and (C)20540261.
T. K.'s work is supported in part by the U. S. National Science Foundation
under Grant NSF-PHY-0757868, and the International Exchange Support
Grants (FY2010) of RIKEN.
T. K. thanks RIKEN for the hospitality extended to him
while a part of this work as carried out.
Numerical calculations are conducted in part on the
RIKEN Super Combined Cluster System (RSCC) and the
RIKEN Integrated Cluster of Clusters (RICC) supercomputing systems.

\end{acknowledgments}


\bibliographystyle{apsrev}
\bibliography{b}

\end{document}